\documentclass{article}

\usepackage[dblblindworkshop,final, preprint]{Styles/neurips_2026}

\usepackage[utf8]{inputenc} 
\usepackage[T1]{fontenc}    
\usepackage{hyperref}       
\usepackage{url}            
\usepackage{booktabs}       
\usepackage{amsmath}
\usepackage{amsfonts}       
\usepackage{amssymb}
\usepackage{graphicx}       
\usepackage{nicefrac}       
\usepackage{microtype}      
\usepackage{xcolor}         
\usepackage{tikz}           
\usetikzlibrary{positioning, arrows.meta, calc}
\usepackage{caption}        
\newcommand{\fitwidth}[1]{%
  \resizebox{\ifdim\width>\linewidth\linewidth\else\width\fi}{!}{#1}}

\title{Scaling subjects in cross-modal alignment: video decoding
with EEG foundation model}

\author{%
  Dung Truong$^1$ \qquad Kuntal Kokate$^1$ \qquad Arnaud Delorme$^{1,2}$ \\
  $^1$UC San Diego, USA \qquad $^2$CNRS, France \\
  \texttt{\{dutruong, kkokate, adelorme\}@ucsd.edu} \\
}
\begin{document}

\maketitle

\begin{abstract}
Naturalistic visual decoding from EEG has long been constrained by cohort size:
existing scaling literature caps out at a few dozen subjects,
leading prior work to conclude that scaling subject cohorts yields minimal performance gains. 
Training a cross-modal EEG--video contrastive encoder on a
cohort larger by more than an order of magnitude, we find the axis is productive
but has an onset. Below $S\approx50$ --- the entirety of the range prior work
occupies --- no model improves meaningfully over an untrained encoder; above it, decoding rises
log-linearly in subject count with no saturation at the top of our ladder, on a
fitted stimulus-feature probe and on fit-free movie-moment retrieval alike, and
the gain transfers to a second, unseen film.
How far the axis carries then depends on initialisation more than on capacity: an
encoder initialised from an EEG foundation model scales ${\sim}1.6\times$ faster
per doubling of the cohort than randomly initialised encoders at two depths, is
the only one still converting subjects into retrieval accuracy at the top of the
ladder, and converges on a fraction of the alignment compute.
Subject scaling on shared naturalistic stimuli is thus an effective and
currently unsaturated frontier for scaling EEG foundation models.
\end{abstract}

\section{Introduction}

Naturalistic visual decoding is a growing area of interest in brain--computer
interfacing research
\citep{song2023nice,liu2024eeg2video,huang2025mindev,cinebrain2025,ergezer2026temporal}.
While large-scale pretraining has been shown to help naturalistic EEG decoding
\citep{banville2026neuralbench}, no work has investigated the effect of data
scaling for movie decoding. Prior scaling work has found that the axis
that plays best to EEG's accessibility advantage --- increasing the number of
subjects --- is precisely the unproductive one
\citep{banville2025scaling}.

The classical stimulus-response literature says the signal that subject scaling
should sharpen is demonstrably there: correlated-component analysis and
inter-subject correlation isolate a component of movie-watching EEG that is
reliably \emph{shared} across viewers of the same film
\citep{dmochowski2012corrca,camenzind2026isc}. The naturalistic-video EEG
literature has simply not been in a position to test that axis, running on
cohorts well below $n\approx50$: 20 subjects in SEED-DV
\citep{liu2024eeg2video,huang2025mindev}, 6 in the only long-form film corpus
\citep{cinebrain2025}, 48 in the largest single cohort prior scaling work uses
\citep{grootswagers2022}. The one corpus large enough to try --- HBN, 2634
participants, most of whom watched the
same four short films \citep{alexander2017hbn,shirazi2024hbneeg} --- has been
used only for clip and task classification \citep{ergezer2026temporal}, never
for stimulus decoding. Using it, we report the largest subject-scaling study for EEG
naturalistic video decoding to date. We annotate every frame with 12 continuous
features and a shot/scene segmentation, and score every checkpoint on both a
fitted feature probe and fit-free cross-modal retrieval, the latter unreported on
continuous film. We run the subject ladder three times: once on an encoder
initialised from a pretrained EEG model and twice on randomly initialised
encoders at two depths --- the \emph{pretrained} arm and the two \emph{random-init}
arms, as we call them throughout. Our contributions:

\begin{enumerate}\itemsep2pt\parskip0pt
\item \textbf{A log-linear subject scaling law, and a floor that locates its
onset} (\S\ref{sec:scaling}): decoding scales log-linearly in subject count
across independent readouts, but only above $S\approx50$; below that no arm is
$1.4\times$ an untrained encoder --- the regime prior scaling work
occupies, making its null and ours the same measurement.

\item \textbf{Initialisation moves the exponent more than capacity does}
(\S\ref{sec:depth}): the pretrained arm scales $1.5$--$1.7\times$ faster per doubling
of the cohort on the probe than either random-init arm, and is the only arm still
converting subjects into retrieval accuracy past 700, where both randomly
initialised depths flatten. Depth shifts the same slopes too, by less, and what it
does to the level flips sign between within- and cross-task.

\item \textbf{Pretraining changes what a fixed budget buys} (\S\ref{sec:budget}):
the pretrained arm's optimum sits inside its schedule at all but one rung while the
random-init optima travel the ladder and pin against their own doubled budget at the
top --- so it reaches a better optimum on at least $2.4\times$ less compute than the
matched-depth random-init arm, and is the arm whose best checkpoint a deployable
stopping rule finds most often.
\end{enumerate}

\section{Method}
\label{sec:method}

\paragraph{Data and targets.} We use 129-channel HBN EEG recordings sampled at 200\,Hz \citep{shirazi2024hbneeg},
from subjects watching two animated movies: \emph{The Present} for training and
\emph{Despicable Me} for cross-task evaluation. EEG is per-channel $z$-scored and
cut into aligned 2\,s windows (Appendix~\ref{app:preproc}).
Out of the 10 public releases we reserve release R6 (108 subjects' recordings, zero overlap with other releases)
as our held-out test set, and subsample training cohorts from the remaining
releases, a pool of 2156 recordings, on the ladder
$S \in \{10, 20, 50, 100, 200, 400, 701, 1000, 1400, 1863\}$ with three nested draws at
each rung. All three arms draw their cohorts with the same seeds,
so arms are compared on identical subjects at every rung.
As prediction targets we annotate the movie frames with
12 continuous features spanning low-level (luminance, contrast, entropy),
mid-level (motion energy, depth, faces) and high-level
(scene naturalness, narrative events) properties (Appendix~\ref{app:features}),
and segment it into shots and scenes,
keeping only those a 2\,s window resolves (Appendix~\ref{app:segmentation}),
which sets the retrieval pools --- and hence chance --- at 101 timepoints, 49 shots and 35 scenes.

\paragraph{Architecture.} We adopt REVE \citep{elouahidi2025reve} as our base architecture
($d_{\text{model}}=512$, depth $22$, $8$ heads, $200$-sample patches with $20$-sample
overlap), which leads competing foundation models on naturalistic decoding
\citep{banville2026neuralbench}; we use its publicly released first-stage weights via
\texttt{braindecode} \citep{schirrmeister2017braindecode}. To vary model scale among
the random-init arms we also train a smaller REVE-like variant, depth $12$ at the
same width, with non-overlapping $400$-sample patches.

\begin{figure}[t]
    \centering
    \includegraphics[width=\linewidth]{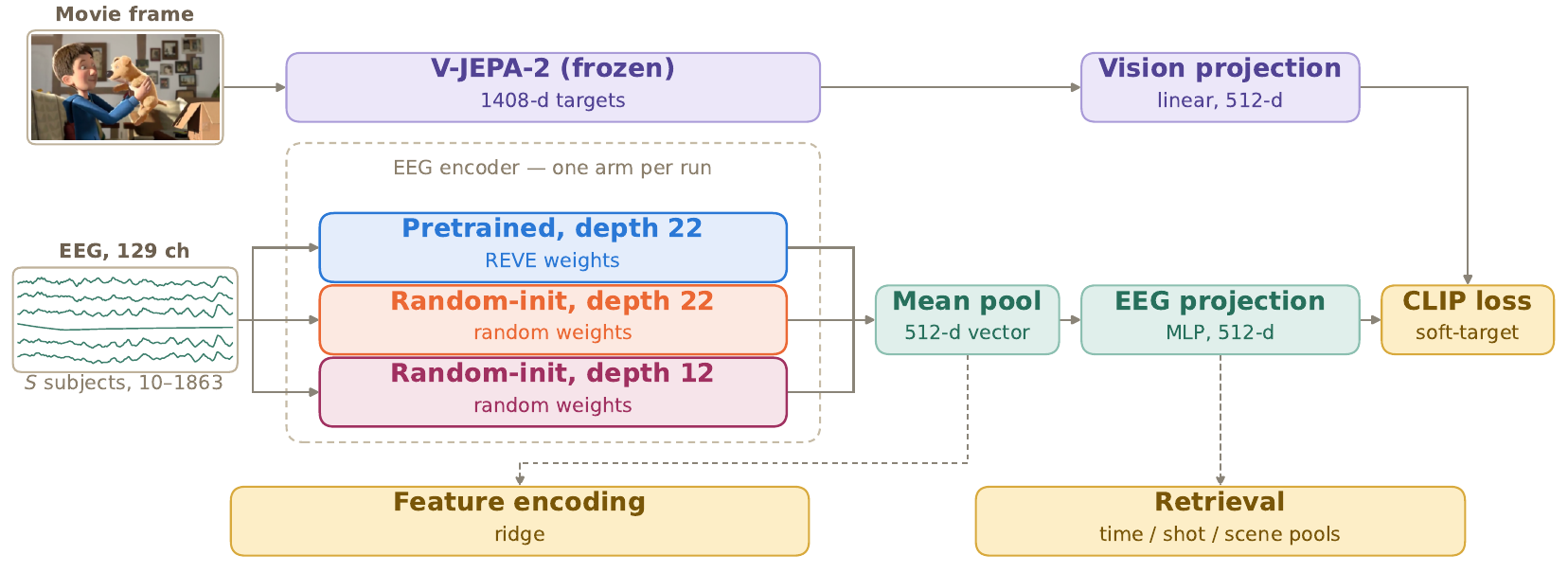}
    \caption{Alignment and readouts. Each training pair is a movie frame and the
    2\,s of EEG recorded against it: a frozen V-JEPA-2 and a linear projection
    place the frame in a 512-d space, and one of three EEG encoders --- differing
    only in initialisation and depth, in the colours they keep throughout the
    paper --- is trained to meet it there under a soft-target CLIP loss. Dashed
    taps mark where the two readouts sit: feature encoding ridge-regresses the
    pooled vector, retrieval scores cosine similarity after the EEG projection.}
    \label{fig:method}
\end{figure}

\paragraph{Objective and training.} EEG-movie alignment is trained using
soft-target CLIP \citep{andonian2022softclip} on mean-centered
movie embedding targets (Figure~\ref{fig:method}, Appendix~\ref{app:centering}). The frozen V-JEPA-2 \citep{assran2025vjepa2} supplying the
vision targets doubles as teacher, its own window-to-window similarities saying
how close those windows should sit in the shared space, so the EEG encoder fits
a graded distance rather than a binary match/mismatch --- a natural fit for
continuous film, where windows seconds apart are genuinely alike. This beat
hard-target CLIP \citep{radford2021clip} and a scene-mask loss (Appendix~\ref{app:softclip});
Appendices~\ref{app:objective} and~\ref{app:batch} give the loss, the optimiser
and the pretrained initialisation in full.

To keep training budget comparable, we fix the number of gradient steps in an epoch and
the total steps per training run, choosing that total from a prior study of where
each model converges (Appendices~\ref{app:probesel} and~\ref{app:compute}).
Below $S=701$
models train for 4400 gradient steps under a cosine schedule annealed over its own budget.
Above $S=701$ the random-init arms were retrained at 8800 steps, the shorter budget
having proved limiting (\S\ref{sec:budget}); since the schedule anneals over the
total, that is a different trajectory rather than a longer one, so convergence is
compared only within the 4400-step runs all arms share. Nothing is stopped
early: checkpoints are written on a shared grid and read afterwards.

\paragraph{Evaluation.} Every 2\,s window is mean-pooled over patch and channel axes
to 512-d, and the two readouts then consume different parts of the model. \emph{Feature encoding}
discards the EEG projection and fits RidgeCV on that vector, scored by Pearson $r$ with a cluster
bootstrap ($B=2000$), isolating representation quality from decoder capacity.
\emph{Retrieval} uses both branches as trained, L2-normalised cosine similarity
in the shared space, e$\to$v top-$\{1,5,10\}$ at all three granularities,
nothing fitted at eval time (v$\to$e is left to
Appendix~\ref{app:retrieval}). \emph{Cross-task} evaluates checkpoints on
the second movie either with the ridge head refit (linear transfer) or entirely
unchanged (zero-shot) (Appendix~\ref{app:crosstask}).
Every cell is 3 nested subject draws, and is read at its own checkpoint rather than a
shared one, so no arm is held to another's stopping point. That checkpoint is chosen
on the draw's cohort complement --- the 293 subjects of the pool it did not train on
--- as the argmax of held-out time-pool retrieval over checkpoints saved every 25
epochs, unsmoothed. Selecting instead on the held-out probe moves the chosen epoch
but not the outcome (Appendix~\ref{app:probesel}). Evaluation is on the test release
only, which no arm and no draw has seen.

\section{Results}

\subsection{Subjects scale, and the flat regime is where prior work sits}
\label{sec:scaling}

\begin{figure}[t]
    \centering
    \includegraphics[width=\linewidth]{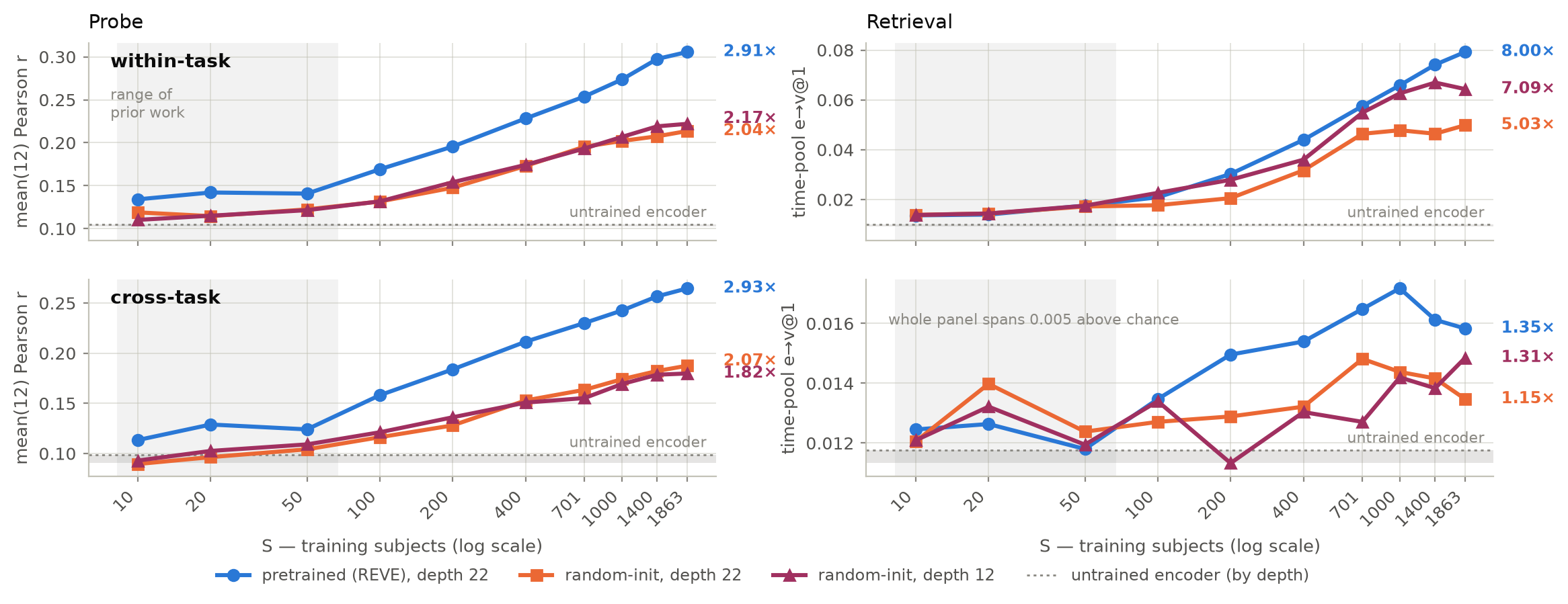}\\[3pt]
    \includegraphics[width=\linewidth]{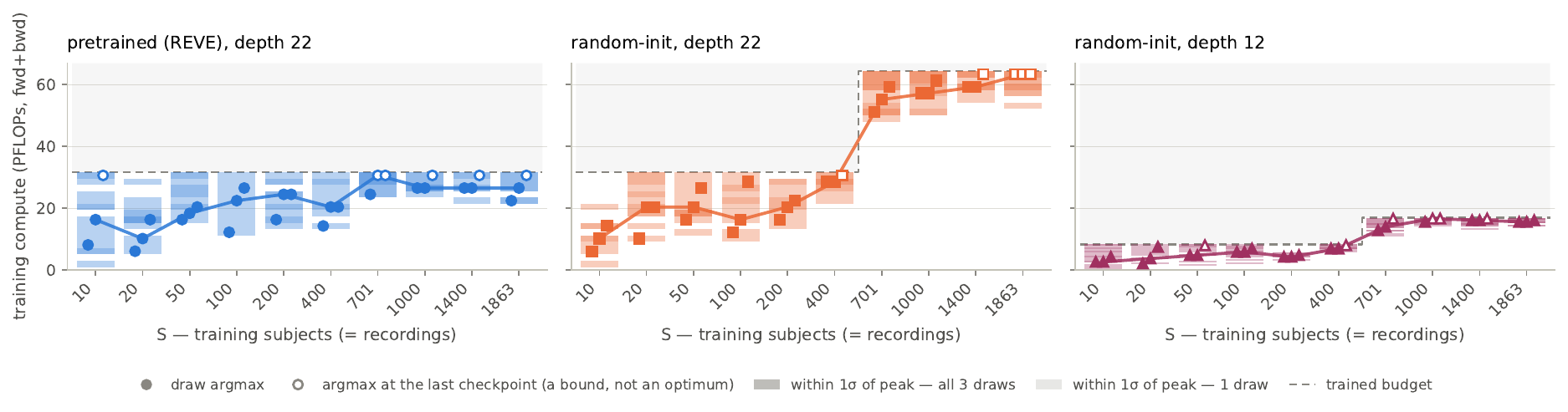}
    \caption{\textbf{Top:} subject scaling on four readouts (rows:
    within- and cross-task; columns: probe and retrieval), test release, each cell
    at its own held-out optimum. Three arms, 3 nested draws per cell; between-draw
    sd (${\approx}0.003$ probe, ${\approx}0.002$ retrieval) is narrower than the
    plotted lines. Dotted rule and band: the random-encoder floor, per depth;
    labels give each arm's terminal score against \emph{its own} floor. Shaded
    column: the range covered by prior work.
    \textbf{Bottom:} where each arm's optimum sits in measured training compute,
    one marker per draw, shaded within one noise $\sigma$ of its peak; hollow
    markers sit on that cell's budget (dashed).}
    \label{fig:scaling}
\end{figure}

\paragraph{Every readout that carries signal rises with subjects.} On the
untouched test release the pretrained arm reaches mean Pearson $r=0.306$ on the
12-feature probe and $8.0\times$ chance on 101-way time-pool retrieval
(Figure~\ref{fig:scaling}, blue). Both fit a log-linear subject law ---
$+0.0244\,r$ and $+0.0093$ top-1 per doubling of the cohort, $R^2=0.94$ and
$0.90$ --- with no sign of saturation. Neither law rests on the top rung:
dropping $S{=}1863$, where three draws share $86\,\%$ of their subjects and so
barely replicate (Limitations), lowers each slope by at most $0.001$ and lowers
$R^2$, and that rung sits \emph{above} the line the nine below it predict.
Appendices~\ref{app:probe},
\ref{app:retrieval} and~\ref{app:fits} give every cell and fit.

\paragraph{The floor is not near zero, and the flat regime is where prior work
sits.} An encoder whose weights are never updated already reaches $r=0.105$ on the
within-task probe --- about a third of our best cell --- so absolute scores here are
uninterpretable without that floor, which every panel therefore carries.
Read against it, the ladder below
$S\approx50$ is barely off the floor: no arm reaches $1.4\times$ an untrained
network on the probe or $2\times$ on retrieval, against $2.0$--$2.9\times$ and
$5.0$--$8.0\times$ at the top. That range is also the one prior work occupies --- the largest cohort of any
modality in the scaling study we contend with \citep{banville2025scaling} is 48
subjects (shaded column) --- so our disagreement with it needs no difference of
method: below $S\approx50$ our own curve moves too little for a design at their
$n$ to resolve, and no image-decoding dataset available to them could have
reached the regime above it.

\paragraph{One readout does not scale at all, and it is our control.} Cross-task
zero-shot retrieval never leaves chance in any arm at any subject count, the whole
range sitting within $0.005$ of its floor (bottom right) --- a negative control
showing the pipeline does not manufacture a subject slope where none exists. The
cross-task \emph{probe} does scale ($2.93\times$ chance at the top), so what fails
to transfer is the EEG projection, not the encoder.

\subsection{A pretrained initialisation lifts the curve and keeps it rising}
\label{sec:init}
\label{sec:depth}

\paragraph{The lift holds on every readout that carries signal.} At matched
depth the pretrained arm leads on both probes at every rung, and on within-task
retrieval from the scaling onset upward; below that onset the arms are
indistinguishable, as \S\ref{sec:scaling} says of the whole floor region.
It is not a head start. What separates the arms is the slope: fitted per
doubling of the cohort, the pretrained arm gains $1.5$--$1.7\times$ as much as
either random-init arm, on the probe and on retrieval alike, so the gap widens
with every rung rather than sitting as a constant offset.
The distinction matters most at the top of the ladder, where the two readouts
part company. Both arms are still converting subjects into probe accuracy there.
Retrieval is where the random-init arms stall: over the top of the ladder
($S{=}701\to1863$) the deeper one adds $+0.003$ top-1 where the pretrained arm
adds $+0.022$, and the shallower one peaks early and turns down. Flattening on
the subject axis is therefore real but not a property of the axis: it is
confined to one readout of the random-init arms, and the pretrained arm does not
share it.

\paragraph{Depth matters under random initialisation, and the sign flips.} Pairing the
two random-init arms draw by draw - each pair trained on identical subjects - the shallower arm wins within-task ($+0.008\,r$ on the probe and $+0.017$ on
retrieval, 9 of 9 pairs) and loses cross-task ($-0.005\,r$, 1 of 9), averaged
over $S\ge1000$ where they separate. However, the advantage peaks at
$S{=}1400$ and narrows at $S{=}1863$ (the shallower arm turns down first)
so it should not be extrapolated past the ladder. And with no pretrained
depth-12 arm, whether that initialisation would erase the gap is untested.
Appendix~\ref{app:depth} gives every cell.

\subsection{Convergence behavior and compute budget}
\label{sec:budget}

\paragraph{Where the optimum sits, and what it costs.} The random-init arms' optimum
travels the length of the ladder --- median epoch 125 at $S=10$ to 775 at
$S=1863$ --- while the pretrained optimum barely moves, sitting at 325 of a 375
budget at the top (Figure~\ref{fig:scaling}, bottom). Two things follow. First,
a fixed epoch count cannot be held across a subject ladder: the budget that
suits the small-$S$ cells truncates the large-$S$ ones, and the apparent
saturation that produces belongs to the protocol, not the subject axis. Hence
each cell is read at its own optimum, and the random-init arms above $S=701$
were retrained at twice the budget --- which did not settle it, since at
$S=1863$ all three depth-22 draws still peak on the last checkpoint of the
longer schedule and so report a bound (hollow markers). Second, the optimum's
location is a cost: $63.3$\,PFLOPs for the depth-22 random-init arm against $26.6$ for
the pretrained arm, which is also the better model on every readout. The
comparison that matters for choosing an architecture, though, is between the
two random-init arms: a depth-22 step costs $3.8\times$ a depth-12 one, so the
shallower arm reaches its optimum on $15.6$\,PFLOPs --- a quarter of the deeper
arm's compute --- while beating it on every within-task readout
(\S\ref{sec:depth}). At this data scale the extra depth buys nothing worth
paying for. The compute that produced the released REVE weights is not counted
in any figure here: already paid, and reusable.

\paragraph{Model selection is reliable for the pretrained arm and unreliable
without it.} Early stopping with patience --- the rule a practitioner would
actually deploy --- selects the best available checkpoint in 12 of 15 high-$S$
cells for the pretrained arm and 8 of 15 for each random-init arm, with patience
set as the same fraction of each arm's own budget. The failures are not near
misses: random-init cells quit at epoch 200 when the best checkpoint lay at
620--680. The arm selects well because it has converged, so without convergence
the stopping rule becomes a second uncontrolled variable
(Appendix~\ref{app:compute}).

\section{Discussion}

Subject count is a productive scaling axis for naturalistic EEG decoding: every
readout carrying signal rises log-linearly and none saturates at the top of our
ladder --- but only above an onset near $S\approx50$, below which no arm
meaningfully beats an untrained encoder. Initialisation matters more than
capacity: the pretrained arm scales faster per doubling, alone keeps converting
subjects into retrieval accuracy at the top, and reaches a better optimum on a
fraction of the compute. Cross-task zero-shot retrieval, our negative control,
never leaves chance. We say nothing about the hours-per-subject law, which a single animated clip
(${\sim}3.4$ minutes per subject) cannot test.

Our disagreement with \citet{banville2025scaling} is a range restriction, not a
contradiction (\S\ref{sec:scaling}). Two design differences predict who should
see an onset: their decoders start from random weights, and our random-init arms
scale at roughly half the pretrained exponent; and their per-subject linear
layers relieve the shared trunk of subject-invariance, where ours makes subjects
the only pressure toward it --- the very differences they blame for their flat
axis.

\paragraph{Limitations.} All of this rests on a single dataset, montage and
protocol, with one training stimulus and one transfer target. Fit and test share
a stimulus on the within-task probe, inflating its absolute $r$; subjects stay
disjoint and the overlap is constant across $S$, so it cannot produce a slope
--- as the cross-task probe, on an unseen film, confirms. Three draws per
cell put the smallest attainable per-step $p$ at $0.10$, so we rest on the
ladder as a whole rather than any one step, and the top rung barely replicates,
each draw there taking $86\,\%$ of the pool --- as in any scaling study
approaching its dataset's size
(Appendix~\ref{app:selection}). Our capacity axis is two depths rather than a
sweep. No arm received its own hyperparameter sweep, so a
residual tuning advantage is possible.

\paragraph{Outlook.} The natural extension is across datasets with heterogeneous
montages but shared stimuli, which REVE's montage-agnostic input supports. Under current data-sharing and
standardisation efforts such as NEMAR \citep{delorme2022nemar}, EEGDash
\citep{eegdash} and HED \citep{robbins2022hed}, subject
pooling is an effective and cheap axis to grow.

\label{probe:refs}

\appendix
\setcounter{table}{0}
\renewcommand{\thetable}{A\arabic{table}}

\newpage
\begin{center}
  {\Large\bfseries Appendix}
\end{center}
\vspace{0.5\baselineskip}

\section{Data and annotation}

\subsection{EEG preprocessing}
\label{app:preproc}

The EEG side of the pipeline runs in two stages. An offline stage is applied
once per release and task and written to disk; a load-time stage runs inside
every training and evaluation job and is what determines the windows a model
actually sees. Separating them matters because normalisation happens in both.

\paragraph{Offline preprocessing.} Recordings shorter than $10$\,s are
discarded. The remainder are resampled from the native $500$\,Hz to $200$\,Hz
and bandpass filtered to $0.5$--$99.5$\,Hz -- a zero-phase FIR filter with a
Hamming window and automatically chosen transition bands, all library defaults
at these cutoffs -- then cast to \texttt{float32}. Each channel is then
$z$-scored using a mean and standard deviation pooled over every time sample of
every recording within one release and task, computed in a single streaming
pass in double precision, and the result is clipped to $\pm15$ standard
deviations. Clipping is the only amplitude-based intervention anywhere in the
pipeline: extreme samples are clamped, never dropped, and no recording or window
is discarded on amplitude grounds. All $129$ channels are carried through
unchanged.

\paragraph{Windowing and stimulus alignment.} At load time a recording is
required to carry both movie-onset and movie-offset event markers, and the
duration they bracket must fall short of the nominal film length by no more
than $1$\,s; recordings failing either check are dropped. This screen is what
sets the realised cohort sizes: the release opt-in yields $1863$ usable
\emph{The Present} recordings, one per subject, but only $1832$ for
\emph{Despicable Me}, and the $1832$ figure is identical across every
cross-task cell, so it is a property of the opt-in rather than per-run
measurement noise. The window grid is
anchored at the movie-onset sample itself and steps in non-overlapping
$400$-sample ($2$\,s) strides, with the bracketed span first trimmed to the
nominal film length so the grid cannot run past the credits and any trailing
partial window dropped. Anchoring on the event rather than on recording start
is what makes window $k$ the same movie time in every subject, and is what
yields the $101$ anchors on \emph{The Present} and $85$ on \emph{Despicable Me}
used throughout. Windows for which no V-JEPA-2 \citep{assran2025vjepa2} target falls inside the $2$\,s
span are dropped rather than zero-filled; because the target grid is uniform,
these are always a trailing trim and never an interior hole. No
visual-processing delay is applied: EEG window $k$ is paired with the stimulus
content of movie seconds $[2k, 2k+2)$.

\paragraph{Normalisation is applied twice.} The per-channel $z$-score quoted in
the main text is the load-time one, recomputed on each clip as it is read: per
channel, over the windows drawn from that recording, which for the single-window
sampling used here is per channel per $2$\,s window. It is computed from the
clip itself, not from stored statistics, and it supersedes the offline global
$z$-score rather than compounding with it. What the offline $z$-score leaves
behind is the frame in which clipping was applied -- the $\pm15$ bound is fixed
in release-and-task units and is permanent -- and unit-scaled files on disk.
This is also the reason the stored offline statistics are never reapplied at
load time. Note that the claim in Appendix~\ref{app:features} that nothing is
standardised between extraction and the probe is scoped to the stimulus feature
chain; it says nothing about the EEG side, where the two normalisations above
both apply.

\paragraph{What is not done.} There is no notch or line-noise filtering, no
re-referencing of any kind, no bad-channel detection or interpolation, no ICA,
ASR or other artifact decomposition, no epoch-level amplitude rejection and no
baseline correction. Two optional load-time transforms exist in the
implementation -- appending a $1$--$8$\,Hz analytic envelope as extra channels,
and a correlated-component spatial projection -- and both are disabled in every
run reported in this paper, which therefore all use the $129$ raw channels.

\subsection{How the 12 probe features are produced}
\label{app:features}

The twelve features are the entire feature-encoding readout: every
\texttt{mean(12)} figure in the main text and in Appendix~\ref{app:probe} is an
average over them. This section specifies each one. Shots and scenes, the other
half of the same annotation pipeline, are in Appendix~\ref{app:segmentation}.

\paragraph{Frame-level extraction.} Every frame of the source video is
annotated at native resolution and native frame rate; nothing is downsampled and
no frames are skipped. \emph{The Present} is $640\times360$ at 23.995\,fps and
\emph{Despicable Me} $624\times336$ at 25\,fps, giving 4877 and 4266 annotated
frames respectively. The \emph{The Present} container declares 4878 frames, of
which the last does not decode, so the two figures differ by one and both appear
in this appendix: shot detection runs on the container and is reported against
4878 (Appendix~\ref{app:segmentation}), while the feature table has one row per
successfully decoded frame and so is 4877 long. Ten
of the twelve are per-frame image statistics or pretrained-model outputs; the
remaining two, \texttt{position\_in\_movie} and
\texttt{narrative\_event\_score}, are derived from the resulting frame table.
Table~\ref{tab:app-features} summarises all twelve.

\paragraph{Low-level statistics.} All five are computed on the full-resolution
frame and lie in $[0,1]$ except entropy. \texttt{luminance\_mean} is the mean
Rec.\,601 luma $0.299R+0.587G+0.114B$ divided by 255, and
\texttt{contrast\_rms} the population standard deviation of that same luma,
likewise divided by 255. \texttt{saturation\_mean} is the mean of the 8-bit HSV
saturation channel over 255. \texttt{edge\_density} is the fraction of pixels
flagged by a Canny detector \citep{canny1986} at hysteresis thresholds
$(100,200)$ on the 0--255 scale, $3\times3$ Sobel aperture and $L_1$ gradient
magnitude. \texttt{entropy} is the Shannon entropy, in bits, of the 256-bin
grayscale intensity histogram, so it is bounded by $[0,8]$ and is a first-order
intensity statistic only -- it carries no spatial term.

\paragraph{Mid-level: motion, depth, faces.} \texttt{motion\_energy} is the
spatial mean of the per-pixel $L_2$ magnitude of dense Farneb\"ack optical flow
\citep{farneback2003} between consecutive grayscale frames (pyramid scale 0.5,
3 levels, window 15, 3 iterations, polynomial order 5, $\sigma=1.2$); the first
frame is 0 by construction. Its units are pixels per frame at native
resolution, normalised by neither frame rate nor frame size, so absolute values
are not commensurate between the two movies -- which is one reason the
cross-task probe is refit rather than transferred in the original scale.
\texttt{depth\_mean} is the spatial mean of a DPT-Large monocular depth map
\citep{ranftl2021dpt} taken on the model's native $384\times384$ output grid
without resampling back to frame resolution; DPT predicts affine-invariant
\emph{inverse relative} depth, so the quantity is unitless and uncalibrated and
larger means nearer. \texttt{n\_faces} and \texttt{face\_area\_frac} come from a
ResNet-10 SSD face detector \citep{liu2016ssd} at $300\times300$ input and
confidence threshold $0.5$; \texttt{face\_area\_frac} is the \emph{sum} of
accepted box areas over frame area, so overlapping boxes double-count and it is
not a union coverage fraction. On this material faces are sparse (mean 0.41 per
frame on TP, at most 2), which bounds how much either can be decoded.

\paragraph{High-level: naturalness, events, position.}
\texttt{scene\_natural\_score} is computed with CLIP ViT-B/32
\citep{radford2021clip} as the raw cosine difference
$\cos(v,t_{\text{nat}})-\cos(v,t_{\text{urb}})$ between the $L_2$-normalised
image embedding and two fixed prompts, ``a photograph of a natural scene with
trees, grass, or water'' and ``a photograph of an urban scene with buildings and
roads''. No softmax and no temperature are applied, which is why its whole range
on \emph{The Present} is only $[-0.065,+0.045]$.
\texttt{position\_in\_movie} is the normalised frame index $t/(N-1)\in[0,1]$
with $N=4877$ (TP) and $4266$ (DM). \texttt{narrative\_event\_score} is a mean of
four whole-movie $z$-scores of absolute first differences,
\begin{align}
\text{narr}[t] &\;=\; \frac{1}{4}\sum_{f\in\mathcal{F}}
  \frac{\big|\Delta f[t]\big| - \mu_{|\Delta f|}}{\sigma_{|\Delta f|}}, \nonumber \\
\mathcal{F} &\;=\; \{\texttt{motion\_energy},\ \texttt{luminance\_mean},\
                \texttt{n\_faces},\ \texttt{depth\_mean}\},
\end{align}
with $\Delta f[0]=0$, the absolute value taken \emph{before} standardisation,
and $\mu,\sigma$ estimated over the whole movie. It is mean-zero by
construction, but its standard deviation is $0.713$ (TP) and $0.732$ (DM) rather
than 1, because the four components are correlated ($r=0.79, 0.75, 0.74, 0.58$
with the composite on TP). It is also strongly right-skewed -- TP median
$-0.21$, max $+8.97$ -- so in practice it is a sparse spike train firing at cuts
and at abrupt motion, lighting or face changes. Two caveats follow and we state
them plainly: this is a \emph{heuristic} event proxy built from low- and
mid-level deltas, not a human narrative annotation, and because its statistics
are per movie it is not comparable across movies in absolute scale. That it is,
alongside \texttt{face\_area\_frac}, the weakest of the twelve at every $S$
(Appendix~\ref{app:probe}) is consistent with how little semantic content the
definition actually carries.

\paragraph{From frames to windows.} Each non-overlapping 2\,s EEG window takes
the feature row of the single frame at its \emph{onset}, at zero assumed
visual-processing delay -- the twelve are not averaged within the window. The
V-JEPA-2 targets are treated differently, mean-pooled over the window's full
2\,s span, and that asymmetry is deliberate: the targets are the training
signal and benefit from averaging, while the probe features are read out and are
easier to interpret as an instantaneous frame statistic. Nothing else is done to
them. No smoothing, clipping, log transform or standardisation is applied
anywhere between extraction and the ridge probe; the internal $z$-scores inside
\texttt{narrative\_event\_score} are the only standardisation in the chain, and
at probe time only the EEG embeddings are standardised while the targets stay
raw.

\begin{table}[t]
\centering
\small
\caption{The 12 probe features. ``Source'' names the extractor: OpenCV
primitives for the low-level statistics and optical flow, DPT-Large for depth,
a ResNet-10 SSD for faces, CLIP ViT-B/32 for naturalness, and the frame table
itself for the two derived features. Range is mean\,$\pm$\,sd over the 4877
annotated frames of \emph{The Present}, with $[\min,\max]$ given instead where
the distribution is too skewed for that to be informative.}
\label{tab:app-features}
\fitwidth{%
\setlength{\tabcolsep}{4pt}
\begin{tabular}{llll}
\toprule
feature & definition & source & range (TP) \\
\midrule
\multicolumn{4}{l}{\emph{low-level}} \\
\texttt{luminance\_mean} & mean Rec.\,601 luma\,/\,255 & OpenCV & $0.482\pm0.160$ \\
\texttt{contrast\_rms} & sd of the same luma\,/\,255 & OpenCV & $0.244\pm0.058$ \\
\texttt{saturation\_mean} & mean HSV $S$ channel\,/\,255 & OpenCV & $0.313\pm0.141$ \\
\texttt{edge\_density} & Canny $(100,200)$ pixel fraction & OpenCV & $0.057\pm0.042$ \\
\texttt{entropy} & 256-bin grey histogram, bits & OpenCV & $7.17\pm1.27$ \\
\midrule
\multicolumn{4}{l}{\emph{mid-level}} \\
\texttt{motion\_energy} & mean Farneb\"ack flow magnitude & OpenCV & $[0,\ 21.0]$ \\
\texttt{depth\_mean} & mean inverse relative depth & DPT-Large & $15.73\pm2.73$ \\
\texttt{n\_faces} & detections at conf.\ $>0.5$ & SSD & $0.41\pm0.50$ (max 2) \\
\texttt{face\_area\_frac} & summed box area\,/\,frame area & SSD & $0.052\pm0.092$ \\
\midrule
\multicolumn{4}{l}{\emph{high-level}} \\
\texttt{scene\_natural\_score} & $\cos(v,t_{\text{nat}})-\cos(v,t_{\text{urb}})$ & CLIP ViT-B/32 & $-0.009\pm0.015$ \\
\texttt{position\_in\_movie} & normalised frame index & derived & $0.500\pm0.289$ \\
\texttt{narrative\_event\_score} & mean $z(|\Delta f|)$ over 4 features & derived & $[-0.31,\ 8.97]$ \\
\bottomrule
\multicolumn{4}{l}{\footnotesize \texttt{motion\_energy} and \texttt{narrative\_event\_score} are strongly right-skewed; sd is $2.40$ and $0.713$.} \\
\end{tabular}}
\end{table}

\subsection{Shot and scene segmentation}
\label{app:segmentation}

\paragraph{Shots.} Shot boundaries are taken from \textbf{TransNetV2}
\citep{soucek2020transnetv2}, which
finds 53 boundaries in \emph{The Present} (4878 container frames at 23.995\,fps;
see Appendix~\ref{app:features} for why the feature table has 4877 rows) and 34
in \emph{Despicable Me} (4266 frames at 25\,fps), giving \textbf{54} and
\textbf{35} shots respectively. We validated that choice against two
independent detectors rather than deriving the segmentation from them.
PySceneDetect finds 53 and 35 boundaries, FFmpeg scene-change detection 56 and
29. At a matching tolerance of 12 frames ($\approx0.5$\,s), pairwise $F_1$ on
\emph{The Present} is $0.962$ (PySceneDetect--TransNetV2), $0.973$
(PySceneDetect--FFmpeg) and $0.982$ (TransNetV2--FFmpeg), and 51 of TransNetV2's
53 boundaries are corroborated by both other detectors. On \emph{Despicable Me}
the corresponding $F_1$ values are $0.986$, $0.906$ and $0.921$, with 29
three-way-agreed boundaries -- FFmpeg is the outlier there, missing six cuts
the other two both find. Shot detection on this material is therefore not the
uncertain step; we use the single detector with the best agreement profile and
keep every boundary it reports.

\paragraph{Scenes.} Shots are grouped into scenes by agglomerative
average-linkage clustering of shot centroids in mean-centered V-JEPA-2 cosine
space, merging at centroid cosine $>0.90$. This yields \textbf{36} scenes for
\emph{The Present} (12 multi-shot scenes, at most 4 shots each) and \textbf{29}
for \emph{Despicable Me} (5 multi-shot scenes, at most 3 each). What the merge
buys is a cleaner grouping geometry (Table~\ref{tab:app-scene}): the number of
groups whose centroid is closer to a neighbouring group's centroid than to its
own spread -- i.e.\ groups that would generate false negatives in a
membership-based contrastive loss -- falls from $12/54$ at shot level to
$4/36$ at scene level, nearest-neighbour top-1 purity rises from $0.825$ to
$0.899$, and same-versus-different-group ROC-AUC from $0.916$ to $0.936$. The
merges are semantically interpretable: alternating angles of one conversation,
and returning locations (one scene collects shots 48--50 and 52, the closing
sequence, with shot 51 correctly left separate -- a case no temporal buffer
could recover).

Choosing $0.90$ trades geometry against embedding fidelity. Looser thresholds
give better-separated groups but a scene-mean embedding that has thrown away
much more of the frame-feature signal; the sweep is in
Table~\ref{tab:app-scene}. We use scene identity only for grouping and
reporting, never as the embedding target: broadcasting shot-means to clips
costs $-0.035$ mean ridge $R^2$ against per-clip targets, while broadcasting
scene-means costs $-0.300$.

\paragraph{Why the retrieval pools are smaller than the segmentations.}
Retrieval pools are the distinct shots and scenes actually sampled by the 2\,s
anchor grid, assigning each window the majority shot within it. For \emph{The
Present} that is 49 of 54 shots and 35 of 36 scenes; for \emph{Despicable Me},
32 of 35 and 26 of 29. The missing entries are all shorter than the 2\,s
window and lose the majority vote in every window they overlap: what decides
it is how a short shot straddles the fixed grid, not duration alone. The five
dropped \emph{The Present} shots run $0.54$--$1.92$\,s and the three dropped
\emph{Despicable Me} shots $1.00$--$1.56$\,s, while a $0.71$\,s shot survives
because it happens to sit near the centre of a window. These are the
$N_{\text{pool}}$ values used
throughout (101/49/35 and 85/32/26), so chance is computed against what the
model was actually asked to discriminate.

\begin{table}[t]
\centering
\small
\caption{Scene merging. Left: grouping geometry at shot level versus at the
chosen scene threshold, measured on mean-centered V-JEPA-2 clip embeddings of
\emph{The Present} (406 clips). ``neg.\ margin'' counts groups sitting closer
to another group's centroid than to their own spread. Right: the threshold
sweep, where ``bcast loss'' is the mean ridge $R^2$ given up by using
scene-mean instead of per-clip embeddings -- the reason we do not use scene
means as targets.}
\label{tab:app-scene}
\begin{tabular}{lcc@{\hskip 2.5em}lccccc}
\toprule
\multicolumn{3}{c}{grouping geometry} & \multicolumn{6}{c}{threshold sweep} \\
\cmidrule(lr){1-3}\cmidrule(lr){4-9}
metric & shot & scene & thr. & scenes & neg.\ margin & AUC & NN@1 & bcast loss \\
& \footnotesize($K{=}54$) & \footnotesize($K{=}36$) & & & & & & \\
\midrule
$\eta^2$              & 0.816 & 0.799 & 0.75 & 19 & 1/19 & 0.950 & 0.933 & $-0.453$ \\
$F$                   & 29.5  & 42.1  & 0.85 & 29 & 3/29 & ---   & 0.914 & $-0.388$ \\
same-group cos        & 0.793 & 0.811 & \textbf{0.90} & \textbf{36} & \textbf{4/36} & \textbf{0.936} & \textbf{0.899} & $\mathbf{-0.300}$ \\
diff-group cos        & $-0.008$ & $-0.025$ & 0.92 & 40 & 8/40 & 0.934 & 0.865 & $-0.268$ \\
$d'$                  & 1.94  & 2.08  & 0.95 & 49 & 10/49 & 0.920 & 0.840 & $-0.106$ \\
ROC-AUC               & 0.916 & 0.936 & & & & & & \\
NN top-1 purity       & 0.825 & 0.899 & & & & & & \\
NN top-5 purity       & 0.696 & 0.800 & & & & & & \\
neg.\ margin          & 12/54 & 4/36  & & & & & & \\
\bottomrule
\end{tabular}
\end{table}

\section{Objective and optimisation}

\subsection{The soft-target objective in detail}
\label{app:objective}

This section states exactly what is optimised. The comparison that selected it
over the alternatives is Appendix~\ref{app:softclip}.

\paragraph{What a batch is.} One training example is a single 2\,s EEG window
from one recording, paired with the V-JEPA-2 embedding of the movie span that
window covers -- frame embeddings mean-pooled over the 2\,s span, then
mean-centered by a single global mean vector computed once over the training
clips (Appendix~\ref{app:centering}). A batch is $N=64$ such examples drawn from
64 independently sampled recordings, so two subjects can contribute the same
movie moment to one batch; such pairs are left in place rather than deduplicated,
and the teacher below assigns them a high target similarity anyway. Shot and
scene identities are not used by this objective: all multi-positive structure
comes from the teacher.

\paragraph{The two projections.} The EEG encoder's tokens are mean-pooled over the
patch and channel axes to a 512-d vector per window and passed through the EEG
projection -- $\mathrm{Linear}(512{\to}512)$, a residual block
(GELU, Linear, dropout $0.5$), LayerNorm -- then $\ell_2$-normalised to
$z^{e}_i$. The vision projection is a learnable $\mathrm{Linear}(1408{\to}512)$ on
the centered target $\hat v_i$, likewise $\ell_2$-normalised, giving $z^{v}_i$.
Logits are $s\,z^{e}_i\!\cdot z^{v}_j$ with a learnable scalar scale $s$
initialised to $1/0.07$ as in CLIP and clamped to $s\le 100$.

\paragraph{Student, teacher, blend.} The student distribution, the teacher
distribution and the blended target are
\begin{align}
p_{ij} &= \mathrm{softmax}_j\!\left(s\, z^{e}_i \cdot z^{v}_j\right), &
q_{ij} &= \mathrm{softmax}_j\!\left(\frac{\langle \hat v_i, \hat v_j\rangle}{\|\hat v_i\|\,\|\hat v_j\|}\Big/ \tau_T\right), &
t_{ij} &= (1-\alpha)\,\mathbf{1}[i=j] + \alpha\, q_{ij},
\end{align}
and the loss is the symmetric cross-entropy of the student against that target,
\begin{equation}
\mathcal{L} = -\tfrac{1}{2N}\left(\textstyle\sum_{ij} t_{ij}\log p_{ij} + \sum_{ij} t_{ji}\log p^{\top}_{ij}\right).
\end{equation}
Two properties matter. First, $q$ is computed on the centered V-JEPA-2 targets
\emph{before} the learnable vision projection, so the teacher geometry is
V-JEPA-2's own and the student cannot move it -- it is a fixed distance
structure over the batch, not a co-adapted one. Second, at $\alpha=0$ the target
is one-hot and the loss reduces exactly to hard-target CLIP
\citep{radford2021clip}, so $\alpha$ interpolates between ``only the paired
window is correct'' and ``correctness is graded by teacher similarity''.

\paragraph{Temporal buffer.} Cross-pairs with $|\Delta t| < 2$\,s in movie time
are masked out of \emph{both} the student and the teacher softmax denominators,
so a near-in-time pair contributes to neither the prediction nor the target;
self-pairs are never masked. Appendix~\ref{app:centering} gives the false-negative
accounting that sets the 2\,s width.

\paragraph{A fixed blend.} In \citet{andonian2022softclip} the teacher is the
student's own running average, so $\alpha$ has to be ramped from zero while the
teacher is still uninformative. Ours is a frozen pretrained video model, equally
informative at step 0 and step 4400, and the schedule accordingly collapses to a
constant $\alpha$.

\paragraph{Settings.} $\alpha=0.5$, $\tau_T=0.05$, buffer $2$\,s, $N=64$, shared
space 512-d, EEG-projection dropout $0.5$, logit-scale init $0.07$. As a training
diagnostic we log the teacher entropy $H(q_i)$ and report $\exp H(q_i)$ as an
effective count of positives per anchor; at $\tau_T=0.05$ it runs at
$\approx2$--$3$ of the 64 batch entries.

\subsection{Why the V-JEPA-2 targets are mean-centered}
\label{app:centering}

Raw V-JEPA-2 clip embeddings carry a dominant shared component, so
\emph{every} pairwise cosine on \emph{The Present} sits in $[0.97, 1.0]$:
same-shot pairs average $0.999\pm0.002$ and different-shot pairs
$0.991\pm0.013$. The discriminative geometry exists -- shot identity explains
$\eta^2=0.89$ of angular embedding variance, $F=54.6$ -- but it is compressed
into a sliver of the cosine range that a temperature-scaled softmax cannot
resolve. Subtracting a single global mean vector, computed once over the
training clips, opens the range: same-shot pairs move to $0.793\pm0.250$,
different-shot pairs to $-0.008\pm0.529$, $d'$ from $0.87$ to $1.94$, and
same-versus-different ROC-AUC from $0.88$ to $0.92$. This is why the teacher
term $q_{ij}$ is computed on centered targets: on raw targets the teacher
softmax is nearly uniform regardless of $\tau_T$, and the soft-target loss
degenerates to the hard-target one.

Centering is free on the readout side. It changes the cosine metric but not
what a linear model can fit, and empirically shot-averaged ridge $R^2$ under
centered and raw embeddings agree to within $0.003$ -- so none of the probe
baselines shift because of it.

We should be clear about what is \emph{not} established. The evidence above is
geometric, measured on the target space directly. There is no controlled
end-to-end ablation of centering at fixed learning rate and epoch budget: the
comparison sometimes quoted for it (a centered run at $+0.0272$ mean $\Delta
R^2$ versus an uncentered one at $+0.0014$) also changes the learning rate
($3{\times}10^{-4}\!\to\!10^{-4}$) and the epoch count ($50\!\to\!499$), so it
cannot carry the attribution. The claim we make is the narrow geometric one.

\paragraph{The 2\,s temporal buffer.} The same analysis sets the buffer width.
Treating a cross-shot pair with centered cosine $>0.5$ as a candidate false
negative, a 2\,s buffer removes 283 of 18{,}445 candidates while sacrificing 25
legitimate negatives ($0.04\,\%$ of them) -- $92\,\%$ of what it removes is a
real false negative, an $11\times$ efficiency. Wider buffers stop paying: at
8\,s purity falls to $56\,\%$ and legitimate negatives are lost roughly
one-for-one. The buffer is also not a solution to false negatives in general,
only to the shot-cut-boundary case: the median candidate false negative is
$\Delta t\approx42$\,s away (95th percentile $\approx138$\,s), i.e.\ visually
similar shots scattered across the film, which is precisely the structure the
soft teacher target handles instead.

\subsection{Choosing the training objective}
\label{app:softclip}

The objective was selected in the regime this paper trains in -- a single
movie, one alignment stage -- on the depth-12 random-init pipeline, against
two alternatives: hard-target CLIP, in which only the paired window is a
positive, and a scene-mask multi-positive loss, which treats every window
sharing an agglomerative scene ID as a positive. Soft-target sits between them,
taking its multi-positive structure from the frozen teacher's own similarities
rather than from a discrete label.

The operating point $\alpha=0.5$, $\tau_T=0.05$ (Appendix~\ref{app:objective}
defines both) comes from a 160-epoch
sensitivity sweep: $\tau_T=0.05$ scored $+0.0316$ mean $\Delta R^2$ against
$+0.0284$ at the default $\tau_T=0.1$ and $+0.0306$ at
$\alpha=0.25,\tau_T=0.1$. The mechanism is teacher sharpness -- at
$\tau_T=0.1$ the teacher distributes mass over $\approx5$--$6$ effective
positives per anchor, at $\tau_T=0.05$ over $\approx2$--$3$, and the sharper
setting wins. Because our teacher is frozen and therefore reliable from step 0,
\citeauthor{andonian2022softclip}'s progressive schedule collapses to a fixed
$\alpha$, as the main text notes.

Table~\ref{tab:app-softclip} gives the head-to-head against vanilla
hard-target CLIP at 400 epochs with three seeds and everything else matched.
The result is regime-dependent, and worth stating precisely because it is
narrower than ``soft-target is better'':

\begin{itemize}
\item \textbf{Single-movie training} -- this paper's regime -- soft-target
wins by $+0.0026$ mean $\Delta R^2$, $2.6\times$ the seed sd, with all three
seeds ordered the same way. It is also $+0.0058$ mean Pearson $r$ ahead on the
R6 test split, but only one seed per arm was run on that protocol and the two
are not the same seed, so we rest the claim on the three-seed $\Delta R^2$
comparison and read the test column as corroboration only.
\item \textbf{Multi-movie training} the two are indistinguishable ($+0.0426$
vs $+0.0423$, gap $+0.0003$ against a seed sd of $0.0005$). Cross-movie
targets are near-orthogonal in the centered space and dominate the softmax
denominator, diluting the within-movie multi-positive mass the teacher
supplies.
\end{itemize}

So the objective is a small, real win at single-movie scale and no win at all
when several movies are pooled. These runs are depth-12 random-init, so the
comparison was not re-run at depth 22 from the REVE weights; we carried the
setting forward on the strength of the single-movie result, which is the
regime every arm in this paper trains in.

\paragraph{Why not the scene mask.} The discrete alternative is the one that
loses on cost rather than on score. At 400 epochs multi-movie all three
objectives land inside $[+0.041,+0.043]$ -- scene-mask $+0.0413$, hard-target
$+0.0423\pm0.0005$, soft-target $+0.0426\pm0.0006$ -- and at the 160-epoch
screen the scene mask ($+0.0299$) sits between the two soft-target settings,
beating the smeared teacher at $\tau_T=0.1$ and losing to the sharp one at
$\tau_T=0.05$. It buys nothing at convergence, and it is the only one of the
three that has to be re-derived per movie: shot boundaries must be detected and
a scene-merge threshold chosen before training can start
(Appendix~\ref{app:segmentation}). Soft-target recovers the same
multi-positive structure from the frozen teacher's similarities with no
annotation at all, which is why it and not the scene mask is the objective
carried into this paper. That segmentation pipeline is still run here -- the
retrieval pools are defined by it -- but nothing in training depends on it.

Two side observations from the same runs corroborate results reported
elsewhere in this appendix. First, the low-level/high-level division of labour
that Appendix~\ref{app:probe} finds on the subject axis shows up as a
checkpoint-level dissociation: a
REVE-pretrained run beat these randomly initialised runs on luminance, contrast,
saturation, entropy and position-in-movie, while the randomly initialised runs beat it
on motion energy, narrative events, edge density, faces, depth and scene
naturalness. Second, v$\to$e retrieval was exactly at chance ($1.0\times$) at
all three granularities for the randomly initialised checkpoint, while a pretrained
one reached $2.0\times$ -- the same magnitude as the better cells in
Table~\ref{tab:app-v2e}, and consistent with v$\to$e being weakly sensitive to
initialisation but not to subject count.

\begin{table}[t]
\centering
\small
\caption{The three candidate objectives, 400 epochs, all other settings matched
(depth 12, random-init, mean-centered per-window targets, 2\,s buffer, lr
$10^{-4}$). $\Delta R^2$ is the 5-fold-CV probe gain over a shape-matched
random encoder, over three seeds except where marked. The test column is
train$\to$test mean Pearson $r$ on R6, and is \emph{not} seed-matched: one seed
per arm was run on it, soft-target at its best validation seed and hard-target
at what turned out to be its weakest, so the gap there is corroboration rather
than a matched margin. \emph{Prior best random-init} and \emph{REVE-initialised}
are reference points on the same test protocol.
$^{\dagger}$single run of an earlier sweep, not one of the three-seed
replicates.}
\label{tab:app-softclip}
\begin{tabular}{llccc}
\toprule
regime & objective & mean $\Delta R^2 \pm$ sd & seeds ordered & test mean $r$ \\
\midrule
single-movie (TP) & hard-target CLIP & $+0.0475\pm0.0010$ & --- & 0.1459 \\
single-movie (TP) & \textbf{soft-target} & $\mathbf{+0.0501\pm0.0010}$ & 3/3 & \textbf{0.1517} \\
multi-movie (TP+DM) & hard-target CLIP & $+0.0423\pm0.0005$ & --- & --- \\
multi-movie (TP+DM) & soft-target & $+0.0426\pm0.0006$ & 2/3 & --- \\
multi-movie (TP+DM) & scene-mask multi-pos.$^{\dagger}$ & $+0.0413$ & --- & --- \\
\midrule
\multicolumn{2}{l}{prior best random-init (500 ep)} & --- & --- & 0.1328 \\
\multicolumn{2}{l}{REVE-initialised (299 ep)} & --- & --- & 0.1715 \\
\midrule
\multicolumn{5}{l}{\emph{160-epoch three-way screen, multi-movie}} \\
\multicolumn{2}{l}{soft-target, $\alpha=0.5$, $\tau_T=0.05$} & $\mathbf{+0.0316}$ & & \\
\multicolumn{2}{l}{soft-target, $\alpha=0.25$, $\tau_T=0.1$} & $+0.0306$ & & \\
\multicolumn{2}{l}{scene-mask multi-positive} & $+0.0299$ & & \\
\multicolumn{2}{l}{soft-target, $\alpha=0.5$, $\tau_T=0.1$} & $+0.0284$ & & \\
\bottomrule
\end{tabular}
\end{table}

\subsection{Optimisation settings}
\label{app:batch}

\paragraph{Budget.} Every cell trains for 400 epochs, and an epoch is a fixed
$703$-window random draw from that cell's cohort rather than a pass over it:
$11$ steps at batch 64, so 4400 optimiser steps for every cell on the ladder,
from $S=10$ to the top rung. The one exception is the pair of random-init arms
above $S=701$, which were retrained at 800 epochs (8800 steps) once the shorter
budget proved binding for them; the pretrained arm needed no such retrain. Since
the schedule anneals over the total, those are different trajectories rather
than longer ones, and Appendix~\ref{app:compute}
measures what the original budget cost. The draw is uniform over the cohort and
independent from epoch to epoch, so coverage stays even across epochs while the
step count is decoupled from $S$. Were an epoch instead a pass over the cohort,
every added subject would also buy added optimisation, and the ladder could not
separate the two. The learning-rate schedule is a function of the epoch count
alone, so it is identical step for step across every cell and both arms.

\paragraph{Optimiser.} Adam without weight decay, peak lr $10^{-4}$, five epochs
of linear warmup then cosine decay to $10^{-6}$, stepped once per epoch;
gradient-norm clipping at $1.0$; single precision throughout, one $44$\,GB GPU
per cell. Checkpoints are written every 25 epochs -- 15 per cell, 31 for the
800-epoch retrains -- which is what makes the per-cell selection of
\S\ref{app:selection} a choice made after training rather than during it.
Nothing is early-stopped, and no validation quantity gates the loop.

\paragraph{Initialisation.} The pretrained arm copies the released REVE encoder
weights and nothing else -- the two projections, the optimiser state and the
schedule all begin fresh. The random-init arm is the
same runs with that copy omitted, so the two arms differ in initialisation and
in nothing else, and cells within an arm differ only in which subjects the
sampler may draw.

\paragraph{Batch size.} Batch 64 is carried over from the random-init development line, where two
comparisons favoured it and none favoured anything larger. At the depth-12
shape and lr $10^{-4}$, batch 128 lost $0.0044$ mean $\Delta R^2$ over 300
epochs with everything else held fixed. In an earlier, shallower configuration,
batch 256 with the learning rate square-root-scaled to $2\times10^{-4}$ reached
$+0.0078$ against $+0.0099$ for batch 64 at lr $10^{-4}$ at 99 epochs.

Two caveats keep this from being a clean batch-size result. Both comparisons
ran under the scene-mask objective rather than the soft-target one, and both
were epoch-matched, so the larger-batch arms also took proportionally fewer
optimizer steps -- exactly the confound the fixed-step budget above removes.
We did not re-run the comparison at matched steps, nor with
gradient accumulation, and nothing below 64 was tried; memory bounded the top
end, with batch 512 exceeding a 141\,GB device and batch 128 a 44\,GB one at
the earlier configuration's footprint. Read the setting as inherited and never
contradicted, not as an optimum we located.

\section{Checkpoint selection and compute}

\subsection{How each cell's checkpoint is chosen}
\label{app:selection}

Every cell is evaluated at its \emph{own} optimum rather than at an epoch shared
across the ladder. The selector is held-out time-pool e$\to$v@1, read at every
saved checkpoint of that cell and taken as a plain argmax -- no smoothing, and
no snapping, since the search ranges over the saved checkpoints themselves. The
held-out set is the draw's cohort complement: with cohorts drawn from a pool of
$2156$, a cell trained on $S$ subjects leaves $2156-S$ of them untouched, and
those recordings are legal selection data for that cell. Nothing in the
selection touches the test release.

Selected epochs run $75$--$375$ for the pretrained arm (median $288$),
$75$--$775$ for the depth-22 random-init arm (median $350$) and $100$--$775$ for
the depth-12 one (median $338$), so no single shared epoch would have been right
for any pair of arms, let alone all three. Six of the pretrained arm's 30 cells
select the last checkpoint available to them, against five and six for the two
random-init arms; those cells report a bound rather than a located optimum and
are marked as such wherever they are plotted. Appendix~\ref{app:compute} gives
the per-rung medians and what the budget costs the cells that pin against it.

\paragraph{The top rung barely replicates, and that is structural.} The three
draws at $S=1863$ are not independent: each takes $1863$ of the pool's $2156$
subjects, so two draws share $86.4\,\%$ of their subjects in expectation. The
between-draw spread at that rung therefore measures something much narrower than
draw variance and should not be read as an error bar comparable to those at
smaller $S$. This is why the top rung is excluded from every fit, and it is not
a defect of this study in particular: any scaling ladder whose last rung
approaches the size of its own dataset has it, and the cost of ignoring it is
concrete. In an earlier sweep, whose top rung \emph{was} the whole pool and so
admitted a single draw, that one cell produced a coherent, five-readout,
apparently-real saturation-then-decline -- which reversed sign as soon as the
pool was enlarged and the same $S$ became drawable three ways. The same step in
$S$ measured $-0.0347\,r$ with one draw and $+0.0125$ with three.

\paragraph{Collapse at initialisation, and the screen for it.} Roughly one cell
in ten trained with this recipe fails silently: the contrastive loss sits at
chance ($\ln 64 = 4.16$) from epoch 1 and never moves, but evaluation still
produces well-formed artifacts with numbers a reader would not flag. A cell like
that is indistinguishable from a genuine low-$S$ result unless the training loss
is inspected, which is why we screen rather than eyeball. The test is relative:
a cell is flagged when its final training loss exceeds $1.30\times$ the median of
its own $S$ group. Relative, because loss scales with cohort size -- $S=10$ sits
near $0.70$ and $S=1863$ near $2.8$ -- so a fixed cutoff would miss the same
failure at low $S$. The threshold separates cleanly on our data: genuine failures
land at $1.67$--$3.09\times$ their group median while the worst healthy cell
reaches $1.19\times$. Flagged cells are re-run at a different seed (2025, 7,
2026 in turn), capped at three attempts so that the procedure cannot become a
search for a preferred answer, and every superseded run is retained on disk with
its artifacts. The observed rate is $10$--$11\,\%$ across two independent sweeps
of 31 and 28 cells run with this recipe. Applied to the random-init arms of
\S\ref{sec:init}, the same screen flagged nothing.

\subsection{A second selector, and what it does not change}
\label{app:probesel}

The selector above is one metric, and a reasonable worry is that the ladder
inherits its noise. We therefore re-selected every cell of all three arms on an
independent signal -- a ridge probe over the 12 features, scored on the same
cohort complement -- and re-read every readout at the epoch it chooses.

The two selectors disagree about the epoch, often sharply. On the pretrained arm
the probe puts $S=100$ at epoch 75 where retrieval puts it at 275, and across
the ladder the probe's choice is the earlier one at 7 of 10 rungs;
Table~\ref{tab:app-optepoch} gives both rows for all three arms. What they do
not disagree about is the outcome:

\begin{center}\small
\begin{tabular}{lcccc}
\toprule
& \multicolumn{2}{c}{mean(12) probe} & \multicolumn{2}{c}{time-pool e$\to$v@1} \\
\cmidrule(lr){2-3}\cmidrule(lr){4-5}
arm & mean $|\Delta|$ & max $|\Delta|$ & mean $|\Delta|$ & max $|\Delta|$ \\
\midrule
pretrained, d22   & 0.0016 & 0.0030 & 0.0034 & 0.0087 \\
random-init, d22  & 0.0026 & 0.0108 & 0.0016 & 0.0075 \\
random-init, d12  & 0.0035 & 0.0125 & 0.0023 & 0.0049 \\
\bottomrule
\end{tabular}
\end{center}

\noindent Over the ten rungs the two protocols differ by a mean $0.0016$ to
$0.0035\,r$ on the probe, against an initialisation gap of $0.015$ to
$0.092\,r$ (Table~\ref{tab:app-depth-probe}) -- between $4\times$ and $50\times$
larger -- and the sign is not consistent, the probe-selected read being higher
at 6, 7 and 8 of the ten rungs on the three arms. On retrieval the ordering is
one-sided in the way it must be, since retrieval is the selecting metric there:
the retrieval-selected read is higher at 9, 8 and 10 of ten rungs. Every claim
in the body -- the log-linear rise, the $S\approx50$ onset, the arm ordering and
the retrieval flattening of the random-init arms -- is unchanged under the
probe selector.

\subsection{Where each arm's optimum sits, and what it costs}
\label{app:compute}

\paragraph{An epoch is a constant unit of work.} Because \texttt{epoch\_size} is
fixed at $703$ windows on every cell, steps per epoch is $11$ regardless of $S$,
and one epoch is the same amount of computation at $S=10$ as at $S=1863$. The
epoch axis is therefore a compute axis up to one constant per architecture. We
measure that constant rather than assuming it, counting matrix-multiply FLOPs
forward and backward (backward taken as $2\times$ forward) through the encoder and
the EEG projection:

\begin{center}\small
\begin{tabular}{lrr}
\toprule
& depth 22 & depth 12 \\
\midrule
encoder parameters      & $69.2$\,M & $37.9$\,M \\
tokens per 2\,s window  & $258$     & $129$ \\
FLOPs per optimiser step & $7.43$\,T & $1.96$\,T \\
FLOPs per epoch          & $81.7$\,T & $21.5$\,T \\
\bottomrule
\end{tabular}
\end{center}

A depth-22 step costs $3.80\times$ a depth-12 one -- more than the $1.83\times$
parameter ratio, because the shallower variant also halves the token count by
using non-overlapping $400$-sample patches. So the two random-init arms, which
are step-matched by design, are \emph{not} compute-matched, and the ladder must be
read in FLOPs before the arms can be compared on cost.

\paragraph{Optimum epoch per cell.} Median over the three draws, at each cell's
own held-out optimum, under both selectors (Table~\ref{tab:app-optepoch}). The
budget is 375 epochs below $S=701$ and 775 at and above it for the random-init
arms; the pretrained arm stayed at 375 throughout.

\begin{table}[h]
\centering\small
\caption{Median optimum epoch across three draws, by arm and selector. Bold marks
a median sitting on that cell's budget ceiling, where the value is a lower bound
rather than a located optimum.}
\label{tab:app-optepoch}
\begin{tabular}{lrrrrrrrrrr}
\toprule
$S$ & 10 & 20 & 50 & 100 & 200 & 400 & 701 & 1000 & 1400 & 1863 \\
\midrule
\multicolumn{11}{l}{\emph{selected on held-out retrieval} (the body's protocol)}\\
pretrained, d22   & 200 & 125 & 225 & 275 & 300 & 250 & \textbf{375} & 325 & 325 & 325 \\
random-init, d22 & 125 & 250 & 250 & 200 & 250 & 350 & 675 & 700 & 725 & \textbf{775} \\
random-init, d12 & 125 & 175 & 225 & 275 & 200 & 325 & 650 & \textbf{775} & 750 & 725 \\
\midrule
\multicolumn{11}{l}{\emph{selected on the held-out probe}}\\
pretrained, d22   & 250 & 125 & 150 &  75 & 100 & 200 & 225 & 275 & 350 & 325 \\
random-init, d22 & 100 &  75 & 150 & 100 & 225 & 300 & 550 & 700 & \textbf{775} & 725 \\
random-init, d12 &  50 & 125 &  75 & 150 & 150 & 275 & 525 & 650 & 675 & \textbf{775} \\
\bottomrule
\end{tabular}
\end{table}

Two things read off this table. The random-init optimum travels the length of the
ladder under either selector, and the pretrained optimum does not -- so no single
fixed epoch can serve both arms, and a protocol that imposes one will report the
random-init arm as having saturated when it has been truncated. And the
random-init arms pin against their ceiling at the top of the ladder even at the
doubled budget, so their terminal cells report bounds. Read nothing off the
columns at $S\le50$: both arms are within ${\sim}10\,\%$ of their peak by the
first saved checkpoint there, so the curve is a plateau and its argmax is placed
by noise rather than by convergence.

\paragraph{What the shorter budget cost.} An argmax at the last checkpoint
cannot distinguish a converged curve from a truncated one, so we measured the
difference instead. The depth-22 random-init arm was trained a second time at
double the step budget (8800 steps, checkpoints to epoch 775), and the shortfall
of the best checkpoint reachable inside the original 4400-step budget is:

\begin{center}\small
\begin{tabular}{lccccc}
\toprule
$S$ & $\le400$ & 701 & 1000 & 1400 & 1863 \\
\midrule
cost of the 4400-step budget ($r$) & $0.0000$ & $0.0014$ & $0.0117$ & $0.0120$ & $0.0156$ \\
\bottomrule
\end{tabular}
\end{center}

\noindent So the budget was binding above $S\approx1000$ and not at all below
$S=400$. Two consequences. The optimum grows with cohort size, which means
larger cohorts need more optimisation and not merely more data. And had we kept
the shorter budget, the random-init arm's high-$S$ points would sit about
$0.012\,r$ low, understating its subject slope and overstating the
initialisation gap at the top of the ladder by roughly that amount -- which is
why those cells were retrained rather than reported as they stood.

\paragraph{Early stopping, in detail.} The rule is patience on the same held-out
retrieval curve, with patience set to $10\,\%$ of each arm's own checkpoint count
so that the unequal budgets are handled by the protocol rather than by a caveat:
two checkpoints ($50$ epochs) for the pretrained arm, three ($75$ epochs) for the
random-init arms. Over the 15 cells per arm at $S\ge400$ it reproduces the argmax
in 12, 8 and 8 cells respectively. The disagreements are not small perturbations
but occasional catastrophic exits, where a temporary plateau exhausts the counter
before the curve resumes climbing: the pretrained arm's three misses cost
$0.0028$--$0.0249$, while the random-init depth-22 arm quits at epoch 200 on two
cells whose optima lie at 625 and 675, costing $0.0401$ and $0.0410$. Mean regret
over the same cells is $0.0000$ for the pretrained arm against $0.0031$ for each
random-init arm.

\paragraph{Scope of these claims.} These are claims about whether an arm finishes
inside its budget, what its optimum costs, and whether a deployable rule can find
it. They are not claims that the pretrained arm trains more smoothly --
between-draw outcome variance is indistinguishable across arms, and its epoch curve
is rougher relative to its own rise -- nor about how quickly the optimum arrives,
since the learning rate anneals over each arm's own budget and the arms were not
given equal ones. Nor is the compute that produced the released REVE weights
counted in any FLOP figure here; that cost is real, but already paid and reusable
across every cell of the ladder.

\section{Full per-cell results}

\subsection{Per-feature probe results, all cells}
\label{app:probe}

Tables~\ref{tab:app-probe-tp}~and~\ref{tab:app-probe-dm} give the full
12-feature breakdown behind the mean(12) numbers quoted in the main text. Each
entry is the mean over the 3 subject draws at that $S$. Between-draw sd on
individual features ranges from $0.0003$ to $0.0212$ across all cells and is
omitted here for width; the \texttt{mean(12)} between-draw spread is drawn as
the shaded band in Figure~\ref{fig:scaling}. The
\texttt{random} row is a fresh-init encoder of identical shape, evaluated
through the identical pipeline.

Two things are visible per feature that the mean hides. First, the top of the
ranking is stable but not fixed. On TP test, \texttt{position\_in\_movie} and
the low-level frame statistics (luminance, contrast, saturation, entropy) hold
the top five places at $S=10$, with \texttt{motion\_energy} 7th; motion then
climbs past four of them to 3rd by $S=1000$. The bottom of the ranking does
not move at all: \texttt{face\_area\_frac} and
\texttt{narrative\_event\_score} are the two weakest features at every $S$ on
that split. Second, the features that gain \emph{most} from subjects are not
the ones that start highest -- \texttt{motion\_energy} goes from $0.1003$ to
$0.3601$ ($3.6\times$) while \texttt{luminance\_mean} goes from $0.1955$ to
$0.3606$ ($1.8\times$). What subject scaling buys is concentrated in exactly
the features a frame-statistics readout is weakest on.

\begin{table}[t]
\centering
\small
\caption{Within-task (\emph{The Present}) probe, per-feature test-split
Pearson $r$, depth-22 pretrained. Columns:
lum = \texttt{luminance\_mean}, con = \texttt{contrast\_rms},
edge = \texttt{edge\_density}, sat = \texttt{saturation\_mean},
ent = \texttt{entropy}, mot = \texttt{motion\_energy},
face = \texttt{n\_faces}, area = \texttt{face\_area\_frac},
dep = \texttt{depth\_mean}, nat = \texttt{scene\_natural\_score},
pos = \texttt{position\_in\_movie}, narr = \texttt{narrative\_event\_score}.}
\label{tab:app-probe-tp}
\fitwidth{%
\setlength{\tabcolsep}{2.5pt}
\begin{tabular}{lccccccccccccc}
\toprule
$S$ & lum & con & edge & sat & ent & mot & face & area & dep & nat & pos & narr & mean(12) \\
\midrule
10 & 0.1955 & 0.1897 & 0.0894 & 0.1875 & 0.1999 & 0.1003 & 0.0868 & 0.0822 & 0.0908 & 0.1048 & 0.2120 & 0.0703 & \textbf{0.1341} \\
20 & 0.2035 & 0.1942 & 0.0941 & 0.1927 & 0.1996 & 0.1114 & 0.0988 & 0.0857 & 0.1039 & 0.1133 & 0.2275 & 0.0805 & \textbf{0.1421} \\
50 & 0.1994 & 0.1844 & 0.1025 & 0.1865 & 0.1974 & 0.1276 & 0.1044 & 0.0844 & 0.0993 & 0.1133 & 0.2103 & 0.0799 & \textbf{0.1408} \\
100 & 0.2261 & 0.2216 & 0.1156 & 0.2160 & 0.2226 & 0.1677 & 0.1291 & 0.1017 & 0.1325 & 0.1419 & 0.2468 & 0.1077 & \textbf{0.1691} \\
200 & 0.2524 & 0.2376 & 0.1521 & 0.2386 & 0.2462 & 0.2126 & 0.1492 & 0.1231 & 0.1549 & 0.1702 & 0.2676 & 0.1424 & \textbf{0.1956} \\
400 & 0.2855 & 0.2653 & 0.1874 & 0.2602 & 0.2672 & 0.2496 & 0.1936 & 0.1627 & 0.1891 & 0.2046 & 0.2951 & 0.1786 & \textbf{0.2282} \\
701 & 0.3107 & 0.2874 & 0.2212 & 0.2875 & 0.2912 & 0.2889 & 0.2160 & 0.1881 & 0.2199 & 0.2313 & 0.3054 & 0.1990 & \textbf{0.2539} \\
1000 & 0.3313 & 0.3013 & 0.2499 & 0.2970 & 0.3007 & 0.3125 & 0.2350 & 0.2125 & 0.2406 & 0.2584 & 0.3326 & 0.2113 & \textbf{0.2736} \\
1400 & 0.3509 & 0.3264 & 0.2718 & 0.3202 & 0.3249 & 0.3434 & 0.2603 & 0.2396 & 0.2604 & 0.2807 & 0.3462 & 0.2458 & \textbf{0.2976} \\
1863 & 0.3606 & 0.3356 & 0.2851 & 0.3334 & 0.3347 & 0.3601 & 0.2654 & 0.2333 & 0.2680 & 0.2876 & 0.3556 & 0.2487 & \textbf{0.3057} \\
random & 0.1402 & 0.1273 & 0.0653 & 0.1352 & 0.1501 & 0.0818 & 0.0930 & 0.0826 & 0.0731 & 0.0958 & 0.1488 & 0.0659 & \textbf{0.1049} \\
\bottomrule
\end{tabular}}
\end{table}

\begin{table}[t]
\centering
\small
\caption{Cross-task (\emph{Despicable Me}) probe, per-feature Pearson $r$:
the same \emph{The Present}-trained depth-22 checkpoints with the RidgeCV head
refit on DM's own 1832-recording train pool. Column abbreviations as in
Table~\ref{tab:app-probe-tp}.}
\label{tab:app-probe-dm}
\fitwidth{%
\setlength{\tabcolsep}{2.5pt}
\begin{tabular}{lccccccccccccc}
\toprule
$S$ & lum & con & edge & sat & ent & mot & face & area & dep & nat & pos & narr & mean(12) \\
\midrule
10 & 0.1287 & 0.0985 & 0.1269 & 0.1103 & 0.1230 & 0.1049 & 0.0897 & 0.1139 & 0.1052 & 0.1171 & 0.1697 & 0.0728 & \textbf{0.1134} \\
20 & 0.1465 & 0.1169 & 0.1403 & 0.1341 & 0.1366 & 0.1167 & 0.1064 & 0.1227 & 0.1135 & 0.1351 & 0.1951 & 0.0826 & \textbf{0.1289} \\
50 & 0.1341 & 0.1102 & 0.1387 & 0.1138 & 0.1340 & 0.1287 & 0.0999 & 0.1190 & 0.1168 & 0.1327 & 0.1819 & 0.0788 & \textbf{0.1240} \\
100 & 0.1737 & 0.1476 & 0.1752 & 0.1564 & 0.1674 & 0.1640 & 0.1244 & 0.1454 & 0.1441 & 0.1699 & 0.2158 & 0.1138 & \textbf{0.1581} \\
200 & 0.2093 & 0.1697 & 0.2125 & 0.1857 & 0.1944 & 0.2012 & 0.1444 & 0.1571 & 0.1512 & 0.1997 & 0.2508 & 0.1305 & \textbf{0.1839} \\
400 & 0.2353 & 0.2049 & 0.2419 & 0.2155 & 0.2243 & 0.2230 & 0.1653 & 0.1802 & 0.1871 & 0.2249 & 0.2848 & 0.1502 & \textbf{0.2114} \\
701 & 0.2590 & 0.2210 & 0.2684 & 0.2380 & 0.2418 & 0.2425 & 0.1844 & 0.1982 & 0.1908 & 0.2485 & 0.2963 & 0.1726 & \textbf{0.2301} \\
1000 & 0.2653 & 0.2294 & 0.2803 & 0.2527 & 0.2501 & 0.2649 & 0.1920 & 0.2023 & 0.2086 & 0.2566 & 0.3189 & 0.1897 & \textbf{0.2425} \\
1400 & 0.2852 & 0.2506 & 0.2957 & 0.2698 & 0.2726 & 0.2791 & 0.1982 & 0.2022 & 0.2293 & 0.2733 & 0.3273 & 0.1968 & \textbf{0.2567} \\
1863 & 0.2886 & 0.2535 & 0.2982 & 0.2726 & 0.2743 & 0.2969 & 0.2109 & 0.2216 & 0.2366 & 0.2781 & 0.3380 & 0.2063 & \textbf{0.2646} \\
random & 0.0673 & 0.0657 & 0.1050 & 0.0710 & 0.0987 & 0.0992 & 0.0887 & 0.1040 & 0.1068 & 0.0907 & 0.1100 & 0.0784 & \textbf{0.0905} \\
\bottomrule
\end{tabular}}
\end{table}

\subsection{Retrieval, all pools and all $K$}
\label{app:retrieval}

Table~\ref{tab:app-retr-tp} gives within-task e$\to$v top-$\{1,5,10\}$ at all
three pool granularities; Table~\ref{tab:app-retr-dm} gives the
cross-task (zero-shot, nothing refit) equivalent. The main text
quotes the top-1 column of the former and uses the latter as the study's
negative control. Chance is $1/N_{\text{pool}}$: $0.0099$ (time, $N=101$),
$0.0204$ (shot, $N=49$), $0.0286$ (scene, $N=35$) for TP, and $0.0118$,
$0.0313$, $0.0385$ for DM.

Three details are only visible at full width. (i) The subject effect is present
at every $K$, not just top-1: on TP test scene pools, top-10 rises $0.357 \to
0.674$ across the ladder alongside top-1's $0.059 \to 0.154$. (ii) The
\texttt{random} baseline is \emph{not} at chance for the coarse pools at high
$K$ -- scene top-10 for a fresh-init encoder is $0.241$ on TP test and $0.469$
on DM test, the latter within $0.04$ of the best trained cell on that movie --
which is why cross-task retrieval must be read against the matched random
encoder rather than against $1/N$, and why the floor is taken at each arm's own
depth. (iii) Cross-task e$\to$v@1 is
$0.012$--$0.017$ against a random baseline of $0.012$ on the time pool across
the entire ladder, which is the flatness the main text reports as the negative
control.

Table~\ref{tab:app-v2e} reports the v$\to$e direction, which fails outright.
It is at or within one pool-step of chance at every $S$ and
every granularity, and shows no trend: $186\times$ more subjects leaves
time-pool v$\to$e@1 exactly where it started, at $0.010$ against a chance of
$0.0099$, and the largest cell anywhere on the ladder is $0.017$ -- one
additional correct query out of 101. Whatever this direction is failing at, it
is not a data-quantity problem, since the axis that moves every other readout by
a factor of two to four moves this one not at all.

\begin{table}[t]
\centering
\small
\caption{Within-task (\emph{The Present}) e$\to$v retrieval accuracy,
depth-22 pretrained, test split. Pools: time ($N=101$), shot ($N=49$),
scene ($N=35$). \texttt{random} is a fresh-init encoder and EEG projection of
identical shape.}
\label{tab:app-retr-tp}
\fitwidth{%
\begin{tabular}{lccc@{\hskip 1.5em}ccc@{\hskip 1.5em}ccc}
\toprule
& \multicolumn{3}{c}{time pool} & \multicolumn{3}{c}{shot pool} & \multicolumn{3}{c}{scene pool} \\
\cmidrule(lr){2-4}\cmidrule(lr){5-7}\cmidrule(lr){8-10}
$S$ & @1 & @5 & @10 & @1 & @5 & @10 & @1 & @5 & @10 \\
\midrule
10 & 0.014 & 0.064 & 0.124 & 0.042 & 0.139 & 0.246 & 0.059 & 0.225 & 0.357 \\
20 & 0.014 & 0.067 & 0.129 & 0.036 & 0.135 & 0.249 & 0.058 & 0.213 & 0.361 \\
50 & 0.018 & 0.074 & 0.139 & 0.040 & 0.148 & 0.265 & 0.061 & 0.228 & 0.376 \\
100 & 0.021 & 0.093 & 0.173 & 0.049 & 0.184 & 0.313 & 0.074 & 0.268 & 0.419 \\
200 & 0.030 & 0.122 & 0.215 & 0.061 & 0.214 & 0.343 & 0.081 & 0.283 & 0.435 \\
400 & 0.044 & 0.157 & 0.258 & 0.088 & 0.281 & 0.426 & 0.120 & 0.369 & 0.532 \\
701 & 0.058 & 0.184 & 0.296 & 0.101 & 0.311 & 0.469 & 0.128 & 0.417 & 0.592 \\
1000 & 0.066 & 0.197 & 0.312 & 0.116 & 0.341 & 0.500 & 0.145 & 0.452 & 0.634 \\
1400 & 0.074 & 0.220 & 0.338 & 0.121 & 0.368 & 0.538 & 0.157 & 0.488 & 0.661 \\
1863 & 0.079 & 0.222 & 0.340 & 0.127 & 0.374 & 0.551 & 0.154 & 0.499 & 0.674 \\
random & 0.010 & 0.050 & 0.099 & 0.014 & 0.077 & 0.166 & 0.026 & 0.120 & 0.241 \\
\bottomrule
\end{tabular}}
\end{table}

\begin{table}[t]
\centering
\small
\caption{Cross-task zero-shot e$\to$v retrieval on \emph{Despicable Me} test:
\emph{The Present}-trained encoder \emph{and} EEG projection, nothing refit.
Pools: time ($N=85$), shot ($N=32$), scene ($N=26$). This is the study's
negative control -- flat in $S$ at every $K$ and every granularity.}
\label{tab:app-retr-dm}
\fitwidth{%
\begin{tabular}{lccc@{\hskip 1.5em}ccc@{\hskip 1.5em}ccc}
\toprule
& \multicolumn{3}{c}{time pool} & \multicolumn{3}{c}{shot pool} & \multicolumn{3}{c}{scene pool} \\
\cmidrule(lr){2-4}\cmidrule(lr){5-7}\cmidrule(lr){8-10}
$S$ & @1 & @5 & @10 & @1 & @5 & @10 & @1 & @5 & @10 \\
\midrule
10 & 0.012 & 0.062 & 0.122 & 0.045 & 0.194 & 0.342 & 0.079 & 0.244 & 0.416 \\
20 & 0.013 & 0.063 & 0.127 & 0.045 & 0.188 & 0.336 & 0.071 & 0.235 & 0.404 \\
50 & 0.012 & 0.065 & 0.128 & 0.042 & 0.181 & 0.325 & 0.064 & 0.229 & 0.395 \\
100 & 0.013 & 0.070 & 0.133 & 0.047 & 0.204 & 0.359 & 0.068 & 0.248 & 0.428 \\
200 & 0.015 & 0.069 & 0.135 & 0.051 & 0.210 & 0.370 & 0.069 & 0.260 & 0.443 \\
400 & 0.015 & 0.073 & 0.141 & 0.050 & 0.224 & 0.392 & 0.072 & 0.269 & 0.459 \\
701 & 0.016 & 0.075 & 0.145 & 0.050 & 0.223 & 0.397 & 0.069 & 0.270 & 0.468 \\
1000 & 0.017 & 0.077 & 0.146 & 0.052 & 0.242 & 0.419 & 0.075 & 0.292 & 0.488 \\
1400 & 0.016 & 0.080 & 0.153 & 0.052 & 0.243 & 0.430 & 0.071 & 0.299 & 0.500 \\
1863 & 0.016 & 0.079 & 0.150 & 0.050 & 0.246 & 0.435 & 0.067 & 0.298 & 0.507 \\
random & 0.012 & 0.060 & 0.118 & 0.035 & 0.177 & 0.390 & 0.029 & 0.229 & 0.469 \\
\bottomrule
\end{tabular}}
\end{table}

\begin{table}[t]
\centering
\small
\caption{v$\to$e top-1 accuracy, within-task, depth-22 pretrained, test split.
Chance is $1/N_{\text{pool}}$, shown in the header. Every cell is
within one retrieved item of chance and nothing trends with $S$.}
\label{tab:app-v2e}
\begin{tabular}{lccc}
\toprule
$S$ & time & shot & scene \\
& \footnotesize(0.0099) & \footnotesize(0.0204) & \footnotesize(0.0286) \\
\midrule
10 & 0.010 & 0.020 & 0.029 \\
20 & 0.013 & 0.027 & 0.038 \\
50 & 0.007 & 0.020 & 0.019 \\
100 & 0.010 & 0.020 & 0.029 \\
200 & 0.010 & 0.020 & 0.029 \\
400 & 0.013 & 0.027 & 0.038 \\
701 & 0.013 & 0.027 & 0.038 \\
1000 & 0.010 & 0.020 & 0.029 \\
1400 & 0.017 & 0.041 & 0.057 \\
1863 & 0.010 & 0.020 & 0.029 \\
random & 0.010 & 0.020 & 0.029 \\
\bottomrule
\end{tabular}
\end{table}

\subsection{Log-linear subject fits, every readout}
\label{app:fits}

Table~\ref{tab:fits} regresses each readout on $\log_2 S$ by ordinary least
squares across the ten ladder points $S=10\ldots1863$, so the slope
reads as a change per doubling of the cohort -- the unit used throughout the
paper. Every point is a 3-draw cell mean.
Alongside the fit we report two distribution-free
statistics that make no linearity assumption at all: ``up/9'' counts how many of
the nine adjacent-$S$ intervals increase, and the sign-test $p$ is the
one-sided probability of that many or more increases under a fair coin. The
log-linear form is a description of the ladder, not a claim about the functional
form of the underlying law -- what the sign test establishes independently is
monotonicity.

Two cautions on reading the slope column. First, slopes are not comparable
across readout \emph{types}: retrieval top-1 accuracy lives on a much smaller
scale than probe Pearson $r$, so its absolute slope is smaller even where the
proportional effect is much larger -- the time pool gains $5.80\times$ over
the ladder against the probe's $2.28\times$. The ``gain'' column is the fair
cross-readout comparison.

Second, the headline slope averages across a regime change. Every pretrained-arm readout
is much flatter below $S\approx50$ than above it, so the pooled fit understates
the law in the range where the law holds -- on the within-task probe, $+0.0321$
with $R^2=0.995$ above the knee against the pooled $+0.0244$ and $R^2=0.935$. The
last three columns give the split. We quote the pooled fit throughout the paper
because it is the conservative number, not because it is the better description;
\S\ref{sec:scaling} discusses the knee.

The split is fixed at $S=50$ in every row so that the columns are comparable
across readouts, but $50$ is not the best breakpoint for every one of them.
Choosing the breakpoint per readout by two-segment residual sum of squares, no
readout of any arm puts time-pool retrieval's at $S=50$: the pretrained arm
prefers $S=100$ (two-segment RSS $2.62\times10^{-5}$ against
$1.24\times10^{-4}$ at $S=50$, upper-segment $R^2$ $0.992$ against $0.970$), and
both random-init arms prefer $S=200$, though with residuals
so close across the three candidates that the choice is not resolved. The probe
splits earlier, but only on the pretrained arm, which prefers $S=50$
($1.57\times10^{-4}$ against $2.97\times10^{-4}$ at $S=100$); both random-init
arms put the probe at $S=100$ as well, so the probe/retrieval ordering below is
a statement about the pretrained arm rather than about all three. The
scene pool does not resolve its candidates on any arm -- on the pretrained arm
$3.53$, $3.29$ and $3.10\times10^{-4}$ at $50$, $100$ and $200$. Fixing the split at $50$ therefore
understates the retrieval rows, which is the direction that does not flatter the
subject claim.

That ordering -- probe first, time pool a doubling later, scene pool not
resolving its candidates at all -- is what readout resolution predicts.
Retrieval top-1 moves in whole retrieved items out of 101, so it cannot register
a change in the underlying representation that the continuous probe already
resolves; it is the same resolution argument that \S\ref{sec:depth} makes on the
arm axis, seen here on the subject axis. Breaking the pretrained arm's time-pool fit
at $100$ rather than $50$ gives $+0.0023$ below and $+0.0144$ above. With three
to four points below any candidate breakpoint, what these data locate is the
\emph{ordering} of the onsets, not their values.

\begin{table}[t]
\centering
\small
\caption{Log-linear fits over the ten ladder points
$S=10\ldots1863$, all three arms with every cell read at its own held-out
retrieval optimum, test split. ``up/9'' counts
the adjacent-$S$ intervals on which the readout increases; the sign-test $p$ is
one-sided. The last three columns split the ladder at the $S\approx50$ knee
(\S\ref{sec:scaling}); the $R^2$ given there is the one for the
$50\!\to\!1863$ fit. Splitting raises the upper-segment $R^2$ above the pooled
one on all four signal-carrying readouts of all three arms, so the knee is not a
property of one arm; the controls stay poorly described either way, as they
should. Cross-task zero-shot retrieval is the negative control.}
\label{tab:fits}
\fitwidth{%
\setlength{\tabcolsep}{4.5pt}
\begin{tabular}{lccccc@{\hskip 1.5em}ccc}
\toprule
& \multicolumn{5}{c}{pooled, $10\!\to\!1863$}
& \multicolumn{3}{c}{split at the knee} \\
\cmidrule(lr){2-6}\cmidrule(lr){7-9}
readout & slope & $R^2$ & gain & up/9 & $p$
        & $10\!\to\!50$ & $50\!\to\!1863$ & $R^2$ \\
\midrule
\multicolumn{9}{l}{\emph{depth-22, initialised from REVE-base}} \\
within-task probe, test & $+0.0244$ & 0.935 & $2.28\times$ & 8/9 & 0.0195 & $+0.0027$ & $+0.0321$ & 0.995 \\
within-task retrieval, scene & $+0.0147$ & 0.900 & $2.59\times$ & 7/9 & 0.0898 & $+0.0007$ & $+0.0199$ & 0.965 \\
within-task retrieval, time  & $+0.0093$ & 0.899 & $5.80\times$ & 9/9 & 0.0020 & $+0.0017$ & $+0.0126$ & 0.970 \\
cross-task probe, test  & $+0.0214$ & 0.962 & $2.33\times$ & 8/9 & 0.0195 & $+0.0042$ & $+0.0265$ & 0.997 \\
\textit{cross-task zero-shot retrieval, scene} & $-0.0004$ & 0.053 & $0.86\times$ & 4/9 & 0.7461 & $-0.0063$ & $+0.0009$ & 0.301 \\
\midrule
\multicolumn{9}{l}{\emph{depth-22, randomly initialised}} \\
within-task probe, test & $+0.0146$ & 0.923 & $1.80\times$ & 8/9 & 0.0195 & $+0.0017$ & $+0.0191$ & 0.983 \\
within-task retrieval, scene & $+0.0091$ & 0.899 & $1.96\times$ & 7/9 & 0.0898 & $+0.0022$ & $+0.0118$ & 0.931 \\
within-task retrieval, time  & $+0.0055$ & 0.866 & $3.60\times$ & 8/9 & 0.0195 & $+0.0015$ & $+0.0074$ & 0.913 \\
cross-task probe, test  & $+0.0138$ & 0.967 & $2.10\times$ & 9/9 & 0.0020 & $+0.0063$ & $+0.0167$ & 0.992 \\
\textit{cross-task zero-shot retrieval, scene} & $+0.0012$ & 0.125 & $1.16\times$ & 5/9 & 0.5000 & $-0.0006$ & $+0.0035$ & 0.414 \\
\midrule
\multicolumn{9}{l}{\emph{depth-12, randomly initialised}} \\
within-task probe, test & $+0.0162$ & 0.946 & $2.01\times$ & 9/9 & 0.0020 & $+0.0049$ & $+0.0207$ & 0.991 \\
within-task retrieval, scene & $+0.0098$ & 0.838 & $1.91\times$ & 7/9 & 0.0898 & $-0.0011$ & $+0.0144$ & 0.953 \\
within-task retrieval, time  & $+0.0078$ & 0.891 & $4.67\times$ & 8/9 & 0.0195 & $+0.0016$ & $+0.0105$ & 0.940 \\
cross-task probe, test  & $+0.0120$ & 0.979 & $1.93\times$ & 9/9 & 0.0020 & $+0.0068$ & $+0.0139$ & 0.989 \\
\textit{cross-task zero-shot retrieval, scene} & $-0.0001$ & 0.002 & $0.88\times$ & 4/9 & 0.7461 & $-0.0056$ & $+0.0019$ & 0.334 \\
\bottomrule
\end{tabular}}
\end{table}

\section{Depth-12 versus depth-22 on every cell}
\label{app:depth}

Tables~\ref{tab:app-depth-probe}~and~\ref{tab:app-depth-retr} give the full
three-arm decomposition behind \S\ref{sec:depth}, on both movies and on both
readouts. Reading
\emph{random-init} against \emph{random-init} isolates depth
($\Delta_{\text{d}}$ = d-22 random-init $-$ d-12 random-init); reading depth-22
\emph{random-init} against depth-22 \emph{pretrained} isolates initialisation
($\Delta_{\text{i}}$ = d-22 pretrained $-$ d-22 random-init). A delta near the draw
spread ($\approx0.003\,r$ on the probe) is not distinguishable from noise.

The depth-22 random-init arm is a deliberate 30-cell replication of the main
sweep with the encoder initialisation left random, and the convergence
screen (Appendix~\ref{app:selection}) found no collapsed cells in it. Its step
budget is matched to the main sweep rather than tuned for training from random
weights, so it should be read as ``random initialisation \emph{at this
budget}'', not as the best achievable randomly initialised model.

The pretrained arm wins all 40 probe cells -- both movies, both random-init
arms, every $S$ -- and the margin grows with $S$ on all of them. The two
random-init arms do not separate anywhere: over the ten ladder points the
mean $|\Delta_{\text{d}}|$ is $0.0045$ within-task and $0.0054$ cross-task,
against a between-draw sd of $0.0029$ and $0.0035$, and the deeper arm is higher
in 3 of 10 cells within-task and 5 of 10 cross-task.

The retrieval table is where the readouts dissociate. Within-task, the pretrained initialisation
advantage appears once $S\gtrsim200$ and then grows ($+0.018$ at $S=1000$,
$+0.029$ at $S=1863$); cross-task \emph{zero-shot} retrieval shows no advantage
for any arm at any $S$ ($+0.003$ at most, at the jitter floor), even though
cross-task \emph{probe} transfer on the same checkpoints shows the full-sized
advantage (at matched depth, $+0.0199$ to $+0.0771$). REVE pretraining improves
the features enough
for a refit head to exploit and not at all for the frozen alignment.

\paragraph{Where the pretrained initialisation buys nothing.} It is not a large-cohort-only
effect: on the probe it is worth a draw-separated $+0.015$ to $+0.033\,r$
already at $S=10$--$20$, on both movies. On retrieval it is worth nothing
measurable until $S\approx200$ ($-0.001$ to $+0.003$ at $S\le100$, with the
two arms' draw bands overlapping at every one of those points), for the
resolution reason above.

Read together, the two comparisons are a sharper version of ``scale the model
with the data'': the thing that has to scale with the data is not the parameter
count but the optimisation already paid for. Both random-init arms, $2\times$
apart in depth, flatten on retrieval at the same place; the pretrained arm at
the larger of those two depths does not. One confound in the depth comparison
remains and does not matter: the depth-12 arm also uses non-overlapping
400-sample patches against depth-22's 200/20, so the matched-init comparison is
depth \emph{and} patchification. Since it comes out null, neither term is doing
anything detectable here.

\begin{table}[t]
\centering
\small
\caption{Probe mean(12) Pearson $r$ for all three arms, test
split, each cell at its own held-out optimum. d12$_{\text{r}}$ = depth-12
randomly initialised, d22$_{\text{r}}$ = depth-22 randomly initialised,
d22$_{\text{p}}$ = depth-22 initialised from REVE-base; all three are means of 3
draws. $\Delta_{\text{d}}$ = d22$_{\text{r}}-$d12$_{\text{r}}$ (depth at
matched initialisation), $\Delta_{\text{i}}$ = d22$_{\text{p}}-$d22$_{\text{r}}$
(initialisation at matched depth).}
\label{tab:app-depth-probe}
\fitwidth{%
\setlength{\tabcolsep}{4.5pt}
\begin{tabular}{lccccc@{\hskip 1.6em}ccccc}
\toprule
& \multicolumn{5}{c}{TP test} & \multicolumn{5}{c}{DM test} \\
\cmidrule(lr){2-6}\cmidrule(lr){7-11}
$S$ & d12$_{\text{r}}$ & d22$_{\text{r}}$ & d22$_{\text{p}}$
    & $\Delta_{\text{d}}$ & $\Delta_{\text{i}}$
    & d12$_{\text{r}}$ & d22$_{\text{r}}$ & d22$_{\text{p}}$
    & $\Delta_{\text{d}}$ & $\Delta_{\text{i}}$ \\
\midrule
10 & 0.1102 & 0.1189 & 0.1341 & $+0.0087$ & $+0.0152$ & 0.0929 & 0.0895 & 0.1134 & $-0.0034$ & $+0.0239$ \\
20 & 0.1150 & 0.1146 & 0.1421 & $-0.0004$ & $+0.0275$ & 0.1026 & 0.0964 & 0.1289 & $-0.0062$ & $+0.0325$ \\
50 & 0.1215 & 0.1224 & 0.1408 & $+0.0009$ & $+0.0184$ & 0.1091 & 0.1042 & 0.1240 & $-0.0049$ & $+0.0199$ \\
100 & 0.1317 & 0.1316 & 0.1691 & $-0.0001$ & $+0.0375$ & 0.1212 & 0.1163 & 0.1581 & $-0.0049$ & $+0.0419$ \\
200 & 0.1540 & 0.1477 & 0.1956 & $-0.0064$ & $+0.0479$ & 0.1361 & 0.1280 & 0.1839 & $-0.0081$ & $+0.0559$ \\
400 & 0.1742 & 0.1730 & 0.2282 & $-0.0013$ & $+0.0553$ & 0.1507 & 0.1529 & 0.2114 & $+0.0021$ & $+0.0586$ \\
701 & 0.1933 & 0.1954 & 0.2539 & $+0.0021$ & $+0.0585$ & 0.1553 & 0.1635 & 0.2301 & $+0.0082$ & $+0.0666$ \\
1000 & 0.2068 & 0.2021 & 0.2736 & $-0.0047$ & $+0.0715$ & 0.1693 & 0.1740 & 0.2425 & $+0.0048$ & $+0.0685$ \\
1400 & 0.2191 & 0.2075 & 0.2976 & $-0.0116$ & $+0.0901$ & 0.1785 & 0.1823 & 0.2567 & $+0.0037$ & $+0.0744$ \\
1863 & 0.2220 & 0.2136 & 0.3057 & $-0.0084$ & $+0.0920$ & 0.1798 & 0.1875 & 0.2646 & $+0.0078$ & $+0.0771$ \\
\bottomrule
\end{tabular}}
\end{table}

\begin{table}[t]
\centering
\small
\caption{Time-pool e$\to$v top-1 accuracy for all three arms, test split, each
cell at its own held-out optimum; columns as in Table~\ref{tab:app-depth-probe}.
Within-task chance is $0.0099$ ($N=101$); cross-task chance is $0.0118$ ($N=85$)
and the matched random encoder scores $0.012$ there -- no arm and no $S$ clears
it.}
\label{tab:app-depth-retr}
\fitwidth{%
\setlength{\tabcolsep}{4.5pt}
\begin{tabular}{lccccc@{\hskip 1.6em}ccccc}
\toprule
& \multicolumn{5}{c}{TP test} & \multicolumn{5}{c}{DM test (zero-shot)} \\
\cmidrule(lr){2-6}\cmidrule(lr){7-11}
$S$ & d12$_{\text{r}}$ & d22$_{\text{r}}$ & d22$_{\text{p}}$
    & $\Delta_{\text{d}}$ & $\Delta_{\text{i}}$
    & d12$_{\text{r}}$ & d22$_{\text{r}}$ & d22$_{\text{p}}$
    & $\Delta_{\text{d}}$ & $\Delta_{\text{i}}$ \\
\midrule
10 & 0.014 & 0.014 & 0.014 & $+0.000$ & $-0.000$ & 0.012 & 0.012 & 0.012 & $-0.000$ & $+0.000$ \\
20 & 0.014 & 0.014 & 0.014 & $+0.000$ & $-0.000$ & 0.013 & 0.014 & 0.013 & $+0.001$ & $-0.001$ \\
50 & 0.018 & 0.017 & 0.018 & $-0.000$ & $+0.000$ & 0.012 & 0.012 & 0.012 & $+0.000$ & $-0.001$ \\
100 & 0.023 & 0.018 & 0.021 & $-0.005$ & $+0.003$ & 0.013 & 0.013 & 0.013 & $-0.001$ & $+0.001$ \\
200 & 0.028 & 0.021 & 0.030 & $-0.007$ & $+0.010$ & 0.011 & 0.013 & 0.015 & $+0.002$ & $+0.002$ \\
400 & 0.036 & 0.032 & 0.044 & $-0.004$ & $+0.012$ & 0.013 & 0.013 & 0.015 & $+0.000$ & $+0.002$ \\
701 & 0.055 & 0.046 & 0.058 & $-0.008$ & $+0.011$ & 0.013 & 0.015 & 0.016 & $+0.002$ & $+0.002$ \\
1000 & 0.063 & 0.048 & 0.066 & $-0.015$ & $+0.018$ & 0.014 & 0.014 & 0.017 & $+0.000$ & $+0.003$ \\
1400 & 0.067 & 0.046 & 0.074 & $-0.021$ & $+0.028$ & 0.014 & 0.014 & 0.016 & $+0.000$ & $+0.002$ \\
1863 & 0.064 & 0.050 & 0.079 & $-0.014$ & $+0.029$ & 0.015 & 0.013 & 0.016 & $-0.001$ & $+0.002$ \\
\bottomrule
\end{tabular}}
\end{table}

\section{Cross-task transfer in detail}
\label{app:crosstask}

\S\ref{sec:scaling} reports that linear transfer to \emph{Despicable Me}
scales with subjects while zero-shot alignment does not. Three details behind
that.

\paragraph{At the bottom of the ladder, transfer is worse than nothing.} In the
depth-12 random-init arm, cross-task probe transfer is \emph{below} its matched
random baseline at $S=10$ ($-0.0060\,r$): a 10-subject encoder carries no
usable information about another movie and slightly damages what a random
projection would have given. The sign flips by $S=20$, and by the top of the
ladder the same arm sits $+0.081$ above that baseline -- a range that runs from
worse-than-random to most of the way to a natively trained model, on a film the
encoder never saw, which is the part of the subject claim that within-movie
memorisation cannot explain.

\paragraph{Percent of native, and which arm it applies to.} The depth-12 arm
carries the only DM-native anchors we trained (two cells trained on
\emph{Despicable Me} with the identical recipe, $S=701$ and $1841$, depth-12
random-init). Measured over the same random baseline, its best transfer cell
reaches $76\,\%$ of the native anchor's margin, and $68\,\%$ of its \emph{own}
within-movie margin. The latter comparison is fair because the
two movies' noise ceilings are near-equal ($0.304$ vs $0.313$, cross-validated
CorrCA \citep{dmochowski2012corrca} with the Schoppe single-trial correction
\citep{schoppe2016measuring}) on a $98.9\,\%$-shared cohort.

\paragraph{The native anchor is not an upper bound above depth 12.} On the
cross-task \emph{probe}, depth-22 pretrained transfer at the top of the ladder
($0.2646$) sits \emph{above} the DM-native anchor ($0.2052$): transferring from
\emph{The Present} with 1863 subjects beats training natively on \emph{Despicable
Me} with 1841. That inversion is not evidence against native training -- the
anchor is a depth-12 random-init model, and it is the same arm-quality term as
\S\ref{sec:depth}, which dominates the native-versus-transfer distinction. The
control that makes this readable is the pair of random-init arms, which are
matched to the anchor's initialisation and do sit below it ($0.1875$ and
$0.1798$ against $0.2052$). On cross-task \emph{retrieval} there is no such
inversion: at scene level the native $S=701$ anchor reaches $0.146$ top-1 where
the transferred depth-12 cells reach $0.053$--$0.068$, below even their own
random floor of $0.106$. The percent-of-native figures above are the depth-12
arm measured against its own matched anchor and stand; they should not be read
onto depth 22, and a depth-22 native cell, which would settle it, does not exist.

\paragraph{Why the zero-shot null is a real failure and not a baseline
artifact.} For the depth-12 arm the transferred model scores \emph{below} a
random encoder at every $K$, while DM-native cells clear the same inflated
baseline -- so the metric can detect genuine alignment when it is present. The
encoder learns something about EEG-during-movie-watching that is not specific to
one film; the EEG projection that maps it into V-JEPA-2 space is specific to
that film. This localises the problem to the projection, which is the cheap part
to retrain.

\end{document}